\documentclass[12pt]{article}

\newcommand{\be}{\begin{equation}}
\newcommand{\ee}{\end{equation}}
\newcommand{\bea}{\begin{eqnarray}}
\newcommand{\eea}{\end{eqnarray}}
\newcommand{\beas}{\begin{eqnarray*}}
\newcommand{\eeas}{\end{eqnarray*}}

\newcommand{\DS}{{\Delta_{\textstyle *}}}

\begin{document}
\begin{titlepage}

\begin{center}

{\Large Locality, bulk equations of motion and the conformal bootstrap}

\vspace{8mm}

\renewcommand\thefootnote{\mbox{$\fnsymbol{footnote}$}}
Daniel Kabat${}^{1}$\footnote{daniel.kabat@lehman.cuny.edu},
Gilad Lifschytz${}^{2}$\footnote{giladl@research.haifa.ac.il}

\vspace{4mm}

${}^1${\small \sl Department of Physics and Astronomy} \\
{\small \sl Lehman College, City University of New York, Bronx NY 10468, USA}

\vspace{2mm}

${}^2${\small \sl Department of Mathematics} \\
{\small \sl Faculty of Natural Science, University of Haifa, Haifa 31905, Israel}

\end{center}

\vspace{8mm}
We develop an approach to construct local bulk operators in a CFT to order $1/N^2$.
Since 4-point functions are not fixed by conformal invariance we use the OPE to categorize possible forms for a bulk operator.
Using previous results on 3-point functions we construct a local bulk operator in each OPE channel.
We then impose the condition that the bulk operators constructed in different channels agree, and hence give rise to a well-defined bulk operator.
We refer to this condition as the ``bulk bootstrap.''

We argue and explicitly show in some examples that the bulk bootstrap leads to some of the same results as the regular conformal bootstrap.
In fact the bulk bootstrap provides an easier way to determine some CFT data, since it does not require knowing the form of the conformal blocks.
This analysis clarifies previous results on the relation between bulk locality and the bootstrap for theories with a $1/N$ expansion, and it identifies a
simple and direct way in which OPE coefficients and anomalous dimensions determine the bulk equations of motion to order $1/N^2$.
\noindent

\end{titlepage}
\setcounter{footnote}{0}
\renewcommand\thefootnote{\mbox{\arabic{footnote}}}

\section{Introduction}
One of the most interesting questions in holographic duality \cite{Maldacena:1997re} is how a bulk spacetime description, on which approximately-local bulk fields propagate, arises from the CFT. One way to attack this question is to construct local bulk observables as CFT operators. 
The CFT operators that reproduce bulk 2-point functions for various spin fields are known \cite{Banks:1998dd, Dobrev:1998md, Bena:1999jv,Hamilton:2005ju, Hamilton:2006az, Hamilton:2006fh, Heemskerk:2012np, Kabat:2012hp, Sarkar:2014dma}, and the corrections needed to reproduce bulk 3-point functions for various spins have been worked out \cite{Kabat:2011rz, Heemskerk:2012mn, Kabat:2012av, Kabat:2013wga, Kabat:2015swa}. Recently there have been interesting suggestions of connections to quantum error correction \cite{Almheiri:2014lwa, Mintun:2015qda} and relations
to twisted Ishibashi states \cite{Verlinde:2015qfa, Miyaji:2015fia, Nakayama:2015mva}. 

In constructing bulk operators one can use two approaches.
\begin{itemize}
\item
Given the bulk equations of motion, one can construct a bulk operator in terms of the CFT which obeys this equation of motion. In this construction one basically builds up the Heisenberg picture operator for a bulk field. While this shows that the construction is possible it does not always illuminate the underlying framework. 
\item
One can proceed by demanding certain bulk locality properties for the bulk operators.  The program then becomes constructing
such entities just from the CFT.  In this approach local bulk equations of motion are an output rather than an input.
\end{itemize}
In \cite{Kabat:2015swa} it was shown that these two approaches are equivalent to ${\cal O}(1/N)$.  The general structure for the 3-point function was understood and was shown to result in the correct bulk equation of motion. In this paper we build on these results and treat a
4-point function. By demanding an appropriate bulk locality condition, we show that we can construct operators directly in the CFT that mimic the behavior of local bulk fields when inserted in a scalar 4-point function.

Since 4-point functions, unlike 3-point functions, are not determined by conformal invariance, we cannot proceed as we did in the 3-point case.  Instead, to characterize the 4-point function we make an OPE expansion in various channels. This reduces the problem to a collection of 3-point functions which we already know how to handle. However we must require that, up to possible field redefinitions, we
obtain the same expression for the bulk field in all channels.
This is a powerful constraint which we call the ``bulk bootstrap.'' It should be equivalent to a subset of the regular bootstrap constraints. Depending on the dimensions of the operators appearing in the initial 4-point function, the bulk bootstrap turns out to determine quite simply either the OPE coefficients of double-trace operators to order $1/N^2$, or their anomalous dimensions to the same order. The bulk bootstrap does not require knowledge of the conformal blocks and is much simpler to solve than the regular bootstrap.

In developing this approach we clarify the connection between locality and the bootstrap seen in \cite{Heemskerk:2009pn, Heemskerk:2010ty} and derived using Mellin amplitudes in \cite{Penedones:2010ue,Fitzpatrick:2011ia}. We show how the OPE coefficients (or anomalous dimensions) to order $1/N^2$ determine the bulk equations of motion to the same order. This approach can also be used in reverse and provides a simple way to deduce CFT data given the bulk equations of motion.
 
We start in section \ref{sect:prel} with a summary of previous results needed for this paper and set normalizations and conventions. In section \ref{sect:4point} we develop the bulk bootstrap approach. In this approach one uses the OPE in various channels to reduce the problem to that of an infinite set of 3-point functions,
then demands that the same result arises from the other channels.  We illustrate this with examples of double-trace scalar exchange in section \ref{sect:4points}. We consider examples in which this approach gives the OPE coefficients for the double-trace operators and examples in which it gives the anomalous dimensions of double-trace operators. It is clear from these examples why the bulk equations of motion are local, at least when only double-trace scalar operators appear in all OPE channels.

In section \ref{sect:alls} we formulate things differently. We demand that the expression for the bulk operator $\phi(z,x)$ computed in one channel gives a local expression inside a 4-point function even away from the OPE limit (and hence in any OPE channel). It turns out to
be far simpler to demand that $(\nabla^2-m^2)\phi(z,x)$ be local inside a 4-point function. This can be implemented simply even in situations where the OPE exchange is only known in one channel.  Remarkably results for the OPE or anomalous dimensions to order $1/N^2$ are recovered quite simply in an infinite family of examples, even if other channels include the exchange of operators with unbounded spins.

In section \ref{sect:allv} we repeat the analysis for vector exchange.
Double-trace exchange involving a vector is treated in section \ref{sect:4pointv}, with the simplifying assumption that in another
channel only scalars are exchanged.  The case of single-trace and double-trace spin one exchange in one channel (with
unbounded spins in other channels) is analyzed in section \ref{sect:allv1}.

\section{Preliminaries\label{sect:prel}}
In this section we review results from previous works that we will need in order to tackle a 4-point function.
For notational purposes we use the following. For any primary scalar ${\cal O}_{\Delta}$ of dimension
$\Delta$, either single trace or multi-trace, we define
\begin{equation}
K{\cal O}(z,x) \equiv \phi^{(0)}(z,x)=\int d^{d} x' K_{\Delta}(z,x|x'){\cal O}_{\Delta}(x')
\label{zerodef}
\end{equation}
where $K_{\Delta}(z,x|x')$ is the smearing function obeying the free field equation of motion (acting on $(z,x)$)
\begin{equation}
(\nabla^2-\Delta(\Delta-d))K_{\Delta}(z,x|x')=0
\end{equation}
We need to define the normalization of the bulk and boundary operators and we choose to have it as
\begin{equation}
\langle K{\cal O}(z,x) {\cal O}(y)\rangle =\frac{\Gamma(\Delta)}{\pi^{d/2}\Gamma(\Delta-\frac{d}{2})}\left ( \frac{z}{(x-y)^{2}+z^2}\right )^{\Delta}
\end{equation}
and \cite{Giddings:1999qu, Klebanov:1999tb, Harlow:2011ke}
\begin{equation}
K{\cal O}(z,x)_{z \rightarrow 0} \rightarrow \frac{z^{\Delta}}{2\Delta-d}{\cal O}(x) 
\end{equation}
This means our normalization for primary operators (single- or multi-trace) is
\begin{equation}
\langle{\cal O}(x) {\cal O}(y)\rangle =\frac{(2\Delta -d)\Gamma(\Delta)}{\pi^{d/2}\Gamma(\Delta-\frac{d}{2})}\frac{1}{(x-y)^{2\Delta}}
\end{equation} 

The results of \cite{Kabat:2015swa} can then be summarized as follows (see also appendix A).
Start with a CFT 3-point function with $\tilde{\gamma}$ as its coefficient.
\begin{equation}
\langle{\cal O}(x) {\cal O}_{1}(y_1) {\cal O}_{2}(y_2)\rangle = \frac{\tilde{\gamma}}{|x-y_1|^{\Delta+\Delta_1-\Delta_2}|x-y_2|^{\Delta+\Delta_2-\Delta_1}|y_1-y_2|^{\Delta_1+\Delta_2-\Delta}}
\label{3pointb}
\end{equation}
We want to promote ${\cal O}(x)$ to a bulk operator inside a 3-point function. The zeroth order expression gives
\begin{equation}
\langle\phi^{(0)}(z,x) {\cal O}_{1}(y_1) {\cal O}_{2}(y_2)\rangle =\frac{1}{(y_1 - y_2)^{2\Delta_1}}
\left[\frac{z}{z^2+(x-y_2)^2}\right]^{\Delta_2-\Delta_1} I(\chi)
\label{3point11}
\end{equation}
Here 
\be
\label{chidef}
\chi=\frac{[(x-y_1)^2+z^2][(x-y_2)^2+z^2]}{(y_1-y_2)^2 z^2}
\ee
and
\begin{equation}
\label{Idef}
I(\chi) = \frac{\tilde{\gamma}}{2\Delta-d} \left(\frac{1}{\chi-1}\right)^\DS F\big(\,\DS,\,\DS - \frac{d}{2} + 1,\, \Delta_i - \frac{d}{2} + 1,\,\frac{1}{1-\chi}\,\big)
\end{equation}
with $\Delta_{\textstyle *}=\frac{1}{2}(\Delta+\Delta_{1}-\Delta_{2})$.
However this expression does not obey bulk causality: it is singular at $\chi = 1$, and in general has a branch cut for $0<\chi<1$,
but these values of $\chi$ correspond to bulk spacelike separation.

To restore bulk causality we
we need to change the definition of the bulk field. The change is given by adding to the zeroth order definition (\ref{zerodef}) a tower of higher-dimension primary double-trace operators built out of the other two operators appearing in the initial 3-point function (\ref{3pointb}).
Let us label the double-trace primary scalar built out of ${\cal O}_1$, ${\cal O}_2$ and $2n$ derivatives (hence with dimension $\Delta_n=\Delta_1+\Delta_2 +2n$) as $({\cal O}_1 {\cal O}_2)_n$. Then we define a new bulk operator
\begin{equation}
\phi(z,x)=\phi^{(0)}(z,x)+\frac{1}{N}\sum_{n=0}^{\infty} a_{n12} K({\cal O}_1 {\cal O}_2)_n (z,x)
\label{bulkimp1}
\end{equation}
where the coefficients $a_{n12}$ are chosen such that  inserting (\ref{bulkimp1}) into (\ref{3point11}) instead of $\phi^{(0)}(z,x)$ will give an expression obeying bulk causality. 

We know from \cite{Kabat:2015swa} that there are infinitely many choices for $a_{n12}$ that satisfy this condition.  They are given by
\begin{equation}
\frac{1}{N}a_{n12} = \lambda \frac{b_{n12}}{M_{n}^{2}-m_{0}^{2}}+\frac{1}{N}\sum_{k=0}^{\infty}\beta_{k}(M_{n}^{2}-m_{0}^2)^{k}b_{n12}
\label{ansol}
\end{equation}
In this expression $m_{0}^{2}=\Delta(\Delta-d)$, $M_n^2 = \Delta_n (\Delta_n - d)$, and the coefficient $\lambda$ is fixed by requiring locality to be
\begin{equation}
\lambda=\frac{\tilde{\gamma}}{\gamma_{F}(\Delta,\Delta_1,\Delta_2)}
\label{lamb}
\end{equation}
Here $\tilde{\gamma}$ is the coefficient of the 3-point function appearing in (\ref{3pointb}) and $\gamma_{F}$ is
\cite{Freedman:1998tz}
\begin{eqnarray}
\gamma_{F}(\Delta_i,\Delta_j,\Delta_k)&=&-\frac{\Gamma[\frac{1}{2}(\Delta_{i}+\Delta_{j}-\Delta_{k})]\Gamma[\frac{1}{2}(\Delta_{i}-\Delta_{j}+\Delta_{k})]\Gamma[\frac{1}{2}(\Delta_{k}+\Delta_{j}-\Delta_{i})]}{2\pi^{d}\Gamma(\Delta_{i}-\frac{d}{2})\Gamma(\Delta_{j}-\frac{d}{2})\Gamma(\Delta_{k}-\frac{d}{2})}\nonumber\\
&\times&\Gamma[\frac{1}{2}(\Delta_{i}+\Delta_{j}+\Delta_{k}-d)].
\label{gijk}
\end{eqnarray}
The coefficients $b_{n12}$ are given by\footnote{These coefficients can be computed using (\ref{bncn}) and the computation of $c_{njk}$ in \cite{Fitzpatrick:2011dm}, or once we identify them as in (\ref{basicbs}), from results in \cite{Fitzpatrick:2010zm}.} (recall $\Delta_n=\Delta_1 +\Delta_2 +2n$)
\begin{eqnarray}
b_{n12}&=&(-1)^{n} \left [ \frac{1}{2\pi^{d/2}}\frac{\Gamma(n+\frac{d}{2})}{\Gamma(\frac{d}{2})\Gamma(n+1)}\frac{\Gamma(\Delta_n-\frac{d}{2}+1)\Gamma(\Delta_n-d+1)}{\Gamma(\Delta_n)\Gamma(\Delta_n-n-d+1)} \right.\nonumber \\
& \times & \left. \frac{\Gamma(\Delta_n-\frac{d}{2}-n)\Gamma(\Delta_1+n)\Gamma(\Delta_2+n)}{\Gamma(\Delta_n -\frac{d}{2})\Gamma(\Delta_1+n+1-\frac{d}{2})\Gamma(\Delta_2+n+1-\frac{d}{2})}\right ]^{1/2}
\label{bnij}
\end{eqnarray}
and have the property that
\begin{equation}
\sum_{n}b_{n12}K_{n} ({\cal O}_{1}{\cal O}_{2})_{n}(z,x)=K{\cal O}_1(z,x)K{\cal O}_2(z,x)\equiv\phi^{(0)}_{1}\phi_{2}^{(0)}(z,x).
\label{basicbs}
\end{equation}
The coefficients $\beta_{k}$ appearing in (\ref{ansol}) are arbitrary and parametrize field redefinitions. They correspond to a field redefinition
\begin{equation}
\phi(z,x) \rightarrow \phi(z,x)+\frac{1}{N}\sum_{k=0}^{\infty}\beta_{k}(\nabla^2-m_{0}^{2})^{k}\phi_{1}\phi_{2}(z,x)
\end{equation}
It follows from these results that the bulk operator (\ref{bulkimp1}) obeys the equation of motion
\begin{equation}
(\nabla^2-m_{0}^{2})\phi_(z,x)=\lambda \phi^{(0)}_{1}\phi_{2}^{(0)}(z,x)+\frac{1}{N}\sum_{k=0}^{\infty}\beta_k (\nabla^2-m_{0}^{2})^{k+1}\phi^{(0)}_{1}\phi_{2}^{(0)}(z,x)
\label{bulkeomb}
\end{equation}
The choice $\beta_k = 0$ corresponds to a minimal cubic vertex in the bulk.

\section{A scalar example\label{sect:4point}}
We start from a scalar 4-point function
\begin{equation}
\langle{\cal O}(x) {\cal O}_{1}(y_1) {\cal O}_{2}(y_2){\cal O}_{3}(y_3)\rangle 
\end{equation}
We want to promote ${\cal O}(x)$ to a bulk operator, so we start as we did in the 3-point function case by looking at  
\begin{equation}
\langle K{\cal O}(z,x) {\cal O}_{1}(y_1) {\cal O}_{2}(y_2){\cal O}_{3}(y_3)\rangle 
\label{4point1}
\end{equation}
But unlike 3-point functions which are fixed by conformal symmetry, the form of a 4-point function is not uniquely determined.  So we do not have a general understanding of the properties of (\ref{4point1}).
Instead the idea will be to use the OPE between two of the other operators ${\cal O}_{i}$, ${\cal O}_{j}$ to reduce the 4-point function to an infinite sum of 3-point functions.

In this section we make a simplifying assumption, that
in the OPE of any pair of operators ${\cal O}_{i}(y_i)$ with ${\cal O}_{j}(y_j)$ the only primary operators which
appear are double-trace scalars.\footnote{We only need to worry about primary operators. We do not need to worry about
global descendants in the OPE expansion since they are
obtained from the primaries by applying a local differential operator.  If the primaries obey causality then so will the descendants. This is true as long as the correlation function with the sum of descendants converges. See the discussion at the end of section \ref{sect:4points}.}
In the bulk this corresponds to assuming that we only have 4-point contact interactions.
For instance we will assume that as $y_2 \rightarrow y_3$ we get
\begin{equation}
{\cal O}_{2}(y_2){\cal O}_{3}(y_3)=\sum_{m}d^{(1)}_{m23}(y_2-y_3)({\cal O}_{2}{\cal O}_{3})_{m}(y_2)+\sum_{k}f^{(1)}_{k}(y_2-y_3)({\cal O}{\cal O}_{1})_{k}(y_2)+\cdots
\label{ope1}
\end{equation}
In counting powers of $N$, one has $d_{m23}^{(1)}\sim 1$ and $f_{k}^{(1)} \sim 1/N^2$.

\subsection{Scalar example --  OPE coefficients\label{sect:4points}}
In this subsection we assume that the dimensions of the operators are such that
\begin{equation}
\frac{1}{2}(\Delta+\Delta_{1}-\Delta_{2}-\Delta_{3})
\end{equation}
is not an  integer. (We will relax this assumption in the next subsection.) Then the first term of (\ref{ope1}) will violate causality when used in (\ref{4point1}), due to a discontinuity across a cut given in (\ref{discont}).  But the last term in (\ref{ope1}) respects causality
to order $1/N^2$, since to this order the combination
\be
\Delta_* = {1 \over 2} \left(\Delta+\Delta_1-\Delta_{k}\right) = - k
\ee
is a non-positive integer, which means the discontinuity across the cut (\ref{discont}) vanishes.
We are thus left with an infinite sum of three point functions each of which is of the form
\begin{equation}
\langle{\cal O}(x) {\cal O}_{1}(y_1) ({\cal O}_{2}{\cal O}_{3})_{m}(y_2)\rangle =\frac{c_{m23}}{|x-y_1|^{\Delta+\Delta_1-\Delta_m}|x-y_2|^{\Delta+\Delta_m-\Delta_1}|y_1-y_2|^{\Delta_1+\Delta_m-\Delta}}
\label{3pointms}
\end{equation}
with $\Delta_m=\Delta_2 +\Delta_3 +2m$.

The coefficient $c_{m23}$ is related to an OPE coefficient $\delta C^{{\cal O}{\cal O}_{1}}_{({\cal O}_{2}{\cal O}_{3})_{m}}$ which is defined as 
\begin{equation}
{\cal O}(x){\cal O}_{1}(y_1)=  \delta C^{{\cal O}{\cal O}_{1}}_{({\cal O}_{2}{\cal O}_{3})_{m}}\frac{({\cal O}_{2}{\cal O}_{3})_{m}}{(x-y_1)^{\Delta+\Delta_{1}-\Delta_{m}}}+\cdots
\end{equation}
Then with our choice of normalization of the 2-point function we get
\begin{equation}
c_{m23}=\delta C^{{\cal O}{\cal O}_{1}}_{({\cal O}_{2}{\cal O}_{3})_{m}}\frac{(2\Delta_m -d)\Gamma(\Delta_{m})}{\pi^{d/2}\Gamma(\Delta_{m}-\frac{d}{2})}
\label{coperel}
\end{equation}

We know how to deal with a 3-point function if we want to promote ${\cal O}(x)$ to a local bulk field $\phi(z,x)$. We need to add an infinite tower of appropriately-smeared higher-dimension scalar primary operators made out of 
${\cal O}_{1}$,  $({\cal O}_{2}{\cal O}_{3})_{m}$ and derivatives. This tower of operators will be labeled by  $({\cal O}_{1}({\cal O}_{2}{\cal O}_{3})_{m})_{n}$ with conformal dimension $\Delta_{n}^{m}=\Delta_1 +\Delta_m +2n$.  A local bulk field is then given by 
\begin{equation}
\phi(z,x)=K{\cal O}(z,x)+\frac{1}{N^2}\sum_{m}\sum_{n}a_{n}^{m}K({\cal O}_{1}({\cal O}_{2}{\cal O}_{3})_{m})_{n}(z,x)
\label{bulkimp4s}
\end{equation}
where the coefficients $a_{n}^{m}$ are chosen so that $\phi$ respects locality to this order.
From the results in the introduction (see (\ref{bulkeomb})) we know the possible solutions for $a_{n}^{m}$. The simplest solution (corresponding to $\beta_{k}=0$) is such that the equation of motion obeyed by the local bulk field becomes
\begin{equation}
(\nabla^2-m_{0}^{2})\phi(z,x)=\frac{1}{N^2}K{\cal O}_{1}(z,x)\sum_{m} \lambda_{m23} K({\cal O}_{2}{\cal O}_{3})_{m}(z,x)
\label{bulkeom4s}
\end{equation}
where from (\ref{lamb}) we have
\begin{equation}
\frac{\lambda_{m23}}{N^2}=\frac{c_{m23}}{\gamma_{F}(\Delta,\Delta_1,\Delta_m)}
\label{lamcerel}
\end{equation}

At this stage it seems there are no constraints on the OPE coefficients, and it looks like one can always satisfy bulk locality. To see how constraints come about, note that we have used the OPE between two of the operators to reduce the 4-point function to a sum of 3-point functions. However the OPE only has a certain radius of convergence, and the expression we got for the bulk field (\ref{bulkimp4s}), (\ref{bulkeom4s})
may not work in other limits. For instance we could take the OPE of two other operators, say ${\cal O}_1$ and ${\cal O}_{2}$. This will  give a different expansion of the 4-point function as a sum of 3-point functions.  We could build a local bulk operator in this channel, which however will produce a different-looking expression for the bulk field. But since the two OPE expansions have an overlapping region of convergence, the two expressions must agree. This is a ``bulk'' version of the conformal bootstrap constraints that any CFT must obey.
So we can take our expression for the bulk operator in one channel and compare it to the expression that we get in a different OPE channel. By demanding that the expressions agree, we get some constraints on the OPE coefficients, which should be consistent with the constraints coming from the usual conformal bootstrap.

It's easier to compare equations of motion, rather than directly compare the bulk operator one gets in each OPE channel. From (\ref{bulkeom4s}) we see that the right-hand side of the equations of motion in the different OPE channels are
\begin{eqnarray}
\nonumber
y_2 \rightarrow y_3  \ \ \ \ \ K{\cal O}_1(z,x)\sum_{m}\lambda_{m23} K_{m} ({\cal O}_{2}{\cal O}_{3})_{m}(z,x)\\
\label{allchannels}
y_1 \rightarrow y_3  \ \ \ \ \ K{\cal O}_2(z,x) \sum_{m}\lambda_{m13} K_{m} ({\cal O}_{1}{\cal O}_{3})_{m}(z,x)\\
\nonumber
y_1 \rightarrow y_2  \ \ \ \ \  K{\cal O}_3(z,x)\sum_{m}\lambda_{m12}K_{m} ({\cal O}_{1}{\cal O}_{2})_{m}(z,x)
\end{eqnarray}
We immediately see that the only solution is 
\begin{equation}
\sum_{m}\lambda_{mij} K_{m} ({\cal O}_{i}{\cal O}_{j})_{m}(z,x)=\lambda K{\cal O}_i(z,x)K{\cal O}_j(z,x)
\end{equation}
For the coefficients $\lambda_{mij}$ we simply get (for any spacetime dimension and any conformal dimension)
\begin{equation}
\lambda_{mij}=\lambda b_{mij}
\label{cex1}
\end{equation}
where $b_{mij}$ are given by (\ref{bnij}).

At this point it becomes clear why the bulk bootstrap gave rise to a local bulk equation of motion.
It can be traced back to form of the equation of motion implied by a CFT 3-point function (\ref{bulkeomb}), which has a local form in
terms of the two other operators.\footnote{From the results of \cite{Kabat:2015swa} this behavior is generic and holds for any CFT 3-point function.}  This result, applied in different OPE channels, translates into a constraint that only a local bulk equation of motion can solve.
This result generalizes and holds whenever one has only double-trace exchange, regardless of the spin of the exchanged operator (for an
explicit treatment see the next section). It explains the observation of \cite{Heemskerk:2009pn, Heemskerk:2010ty}, of a connection between
the conformal bootstrap and bulk locality.

The case treated above corresponds to a quartic bulk coupling ${\cal L}_{\rm int }=\frac{\lambda}{N^2} \phi \phi_1 \phi_2 \phi_3$.
The computation of $\lambda_{mij}$ from the bulk bootstrap is a computation of an OPE coefficient, which when all ${\cal O}_{i}$ are distinct (so
that disconnected diagrams do not contribute) is a number of order $1/N^2$.
For this simple case we get from (\ref{coperel}), (\ref{cex1}) the result
\begin{equation}
\delta C^{{\cal O}{\cal O}_{1}}_{({\cal O}_{2}{\cal O}_{3})_{m}}=\frac{\lambda}{N^2}\frac{\pi^{d/2}\gamma_{F}(\Delta,\Delta_1, \Delta_m) \Gamma(\Delta_m-\frac{d}{2})}{(2\Delta_m-d)\Gamma(\Delta_m)}b_{m23}
\label{decoefsim}
\end{equation}
The case $\Delta=\Delta_1$ and $\Delta_2=\Delta_3$ was considered in \cite{Fitzpatrick:2011dm}.  Our result gives for this case
\begin{eqnarray}
\delta C^{{\cal O}{\cal O}_{1}}_{({\cal O}_{2}{\cal O}_{3})_{m}}& = & \frac{(-1)^{m+1}\lambda}{N^2}\frac{\Gamma^{3}(\Delta_2+m)}{4\pi^{d/2}\Gamma^2(\Delta-\frac{d}{2})}
\frac{\Gamma(\Delta-\Delta_2-m)\Gamma(\Delta+\Delta_2+m-\frac{d}{2})}{\Gamma(\Delta_2+m+1-\frac{d}{2})\Gamma(2\Delta_2+2m)} \\
\nonumber
&\times&  \left [  \frac{\Gamma(m+\frac{d}{2})\Gamma(2\Delta_{2}+2m-d+1)\Gamma(2\Delta_2+m-\frac{d}{2})}{\pi^{d/2}(2\Delta_{m}-d)\Gamma(d/2)\Gamma(m+1)\Gamma(2\Delta_2+m-d+1 )\Gamma(2\Delta_2+2m) } \right ]^{1/2}
\end{eqnarray}
which agrees (up to an overall factor) with \cite{Fitzpatrick:2011dm}  after the different normalizations of the 2-point functions are taken into account.

One could consider a more complicated situation, where one allows for a field redefinition (the parameters $\beta_{k}$ appearing in (\ref{bulkeomb})).
The field redefinition will change the equation of motion, but the condition $\lambda_{mij}=\lambda b_{mij}$ which fixes the CFT data will stay the same.
This means there's no change in the OPE coefficients, as one would expect.

This example shows how CFT data -- in this case OPE coefficients at order $1/N^2$ -- directly builds up bulk interactions to the same order.  We will see more examples of this below.

Before moving on to another example we would like to go back and better understand two issues. First, when using the OPE expansion we have neglected all descendant operators.  Is this really justified?  Second, from the bulk operator point of view, what does it mean to say there is an overlapping region of convergence of the OPE expansion?  In fact these two questions are linked.
To see this we use the recent observation \cite{daCunha:2016crm,Czech:2016xec} that in the OPE of two boundary operators (at boundary points $y_1$ and $y_2$), the contribution of a particular conformal family (a primary and all of its descendants) can be represented by integrating
the bulk field we called $\phi^{(0)}(z,x)$ over a bulk geodesic connecting $y_1$ and $y_2$. From this point of view the singularity in (\ref{3point11}) originates from the coincident or light-cone singularity of the bulk two-point function $\langle \phi^{(0)} \phi^{(0)} \rangle$. From the CFT point of view, this singularity can be understood as coming from the sum over the infinite series of descendants which appear in the OPE.  This sum
can diverge as $y_1$ and $y_2$ are changed and must be defined outside its region of convergence by analytic continuation. Thus the approximation of neglecting descendants in a particular OPE channel means that we actually only constructed a local bulk operator outside the causal development of the geodesic which connects the two boundary points that participate in the OPE expansion. But when all OPE channels are considered there is a common bulk region where the bulk operator should be local in all channels.  This is the region where we require the different representations to agree. But once they agree in this region, it is clear from the different-looking but equivalent representations in the different OPE channels that the bulk operator
we have constructed  is local everywhere in the bulk.

\subsection{Scalar example -- anomalous dimensions\label{sect:4pointan}}
We saw in the previous subsection that the bulk equations of motion were fixed by the $1/N^2$ corrections to the OPE coefficients. This is true as long as there is no special relationship between the conformal dimensions of the operators.
In this subsection we will deal with the case that there is.
As a simple example we start with the situation that ($k_1$ and $k_2$ are any integers)
\begin{eqnarray}
\Delta+\Delta_1=\Delta_2+\Delta_3 \nonumber\\
\label{anomcond1}
\Delta+\Delta_2 \neq \Delta_1+\Delta_3 +2k_1 \\
\Delta+\Delta_3 \neq \Delta_1+\Delta_2 +2k_2 \nonumber
\end{eqnarray}
The first condition means that to ${\cal O}(1/N)$ the operators $({\cal O}{\cal O}_1)_{m}$ and 
$({\cal O}_2{\cal O}_3)_{m}$ have the same conformal dimension and can mix (see also the discussion in \cite{D'Hoker:1999jp, Heemskerk:2010ty}).
To see where this issue arises consider what happens if we follow the same procedure as before. Naively it seems to ${\cal O}(1/N^2)$ there won't be a causality-violating discontinuity in a 3-point function with either
$({\cal O}{\cal O}_1)_{m}$ or $({\cal O}_2{\cal O}_3)_{m}$, since (\ref{discont}) vanishes for this case.\footnote{$\Delta_*$ is either zero or a negative
integer.} But this should not be the case, since tuning the masses of the bulk fields (the conformal dimensions of the corresponding operators) cannot turn off
interactions. Another way to see that something non-trivial is happening is to note that in the result (\ref{decoefsim}) $\gamma_{F}$ diverges
when (\ref{anomcond1}) is satisfied.

The correct treatment is to note that while $({\cal O}{\cal O}_1)_{m}$ and $({\cal O}_2{\cal O}_3)_{m}$ are primary operators at leading order in $1/N$,
at ${\cal O}(1/N^2)$ they acquire an anomalous dimension and mix, so by themselves they are not primary operators. As such they are
not the operators that appear in the OPE. Instead, in this simple case where only two operators mix at each level and since the dilation operator acts as a symmetric real matrix in this subspace \cite{D'Hoker:1999jp}, the
primary operators are
\begin{eqnarray}
({\cal O}_2{\cal O}_3)^{+}_{m}=\frac{1}{\sqrt{2}}[({\cal O}_2{\cal O}_3)_{m}+({\cal O}{\cal O}_1)_{m}]\nonumber\\
({\cal O}_2{\cal O}_3)^{-}_{m}=\frac{1}{\sqrt{2}}[({\cal O}_2{\cal O}_3)_{m}-({\cal O}{\cal O}_1)_{m}]
\end{eqnarray}
with dimensions $\Delta_m +\delta \Delta_m$ and $\Delta_m -\delta \Delta_m$ respectively.
So in the OPE channel $y_2 \rightarrow y_3$ there will appear now
 \begin{equation}
{\cal O}_{2}(y_2){\cal O}_{3}(y_3)=\sum_{m}d^{(+)}_{m23}(y_2-y_3)({\cal O}_{2}{\cal O}_{3})^{+}_{m}(y_2)+\sum_{m}d^{(-)}_{m23}(y_2-y_3)({\cal O}_{2}{\cal O}_{3})^{-}_{m}(y_2)
\label{ope1a}
\end{equation}
where to leading order in $1/N$ we have $d^{(+)}_{m23}(y_2-y_3)=d^{(-)}_{m23}(y_2-y_3)$.
Inserting this into the 4-point function results in an infinite set of 3-point functions whose coefficients (see (\ref{3pointms})) we will label by $c_{m23}^{+}$ and $c_{m23}^{-}$ respectively.
We can now make these 3-point functions local as before, by adding higher dimension operators to get an equation of motion
\begin{eqnarray}
(\nabla^2-m_{0}^{2})\phi(z,x)&=&\frac{1}{N^2}\sum_{m} \lambda^{+}_{m23} K{\cal O}_{1}(z,x)K({\cal O}_{2}{\cal O}_{3})^{+}_{m}(z,x)\nonumber\\
&+&\frac{1}{N^2}\sum_{m} \lambda^{-}_{m23} K{\cal O}_{1}(z,x)K({\cal O}_{2}{\cal O}_{3})^{-}_{m}(z,x)
\label{bulkeom4sa}
\end{eqnarray}
Comparing to the other two channels that do not change in (\ref{allchannels}), we see that
\begin{equation}
\lambda^{+}_{m23} = \lambda^{-}_{m23}=\frac{\lambda}{\sqrt{2}} b_{m23},
\label{bostpms}
\end{equation}
and as before (see (\ref{lamcerel})) the relationship is
\begin{equation}
\frac{\lambda^{\pm}_{m23}}{N^2}=\frac{c_{m23}^{\pm}}{\gamma_{F}(\Delta,\Delta_1,\Delta_{m} \pm\delta \Delta_{m})}.
\label{lamcerel+}
\end{equation}
However now the CFT data that we get is different, since to leading order in $1/N$ we have
\begin{equation}
c_{m23}^{+}=-c_{m23}^{-}=\frac{1}{\sqrt{2}}C^{{\cal O}{\cal O}_1}_{({\cal O}{\cal O}_1)_{m}}\frac{(2\Delta_{m}-d)\Gamma(\Delta_m)}{\pi^{d/2}\Gamma(\Delta_m -\frac{d}{2})}
\end{equation}
where $C^{{\cal O}{\cal O}_1}_{({\cal O}{\cal O}_1)_{m}}$ is the ${\cal O}(N^0)$ OPE coefficient defined by
\begin{equation}
{\cal O}(x){\cal O}_{1}(y_1)\sim  C^{{\cal O}{\cal O}_{1}}_{({\cal O}{\cal O}_{1})_{m}}\frac{({\cal O}{\cal O}_{1})_{m}}{(x-y_1)^{\Delta+\Delta_{1}-\Delta_{m}}}+\cdots.
\end{equation}

In fact (\ref{bostpms}), (\ref{lamcerel+}) determine the anomalous dimensions of the double-trace operators.
Using $1/\Gamma(-m-\frac{\delta \Delta_m}{2})=(-1)^{m+1}\Gamma(m+1)\frac{\delta \Delta_m}{2}+O((\delta \Delta_m)^2)$, one gets the relationship
\begin{eqnarray}
\frac{\lambda}{N^2}b_{m23}=\frac{\pi^{d/2}(-1)^m C^{{\cal O}{\cal O}_{1}}_{({\cal O}{\cal O}_{1})_{m}}(2\Delta_m-d)\Gamma(\Delta_m)\Gamma(\Delta-\frac{d}{2})\Gamma(\Delta_1-\frac{d}{2})\Gamma(m+1)}{\Gamma(\Delta+m)\Gamma(\Delta_1 +m)\Gamma(\Delta+\Delta_1+m-\frac{d}{2})}\delta \Delta_{m}\nonumber\\
\end{eqnarray}
Using the known result \cite{Fitzpatrick:2011dm}, and taking into account the different normalizations, we have 
\begin{eqnarray}
C^{{\cal O}{\cal O}_{1}}_{({\cal O}{\cal O}_{1})_{m}}&=&\left [\frac{4\Gamma(\frac{d}{2})}{\pi^{\frac{d}{2}}\Gamma^{2}(\Delta-\frac{d}{2})\Gamma^{2}(\Delta_1-\frac{d}{2})}\right ]^{\frac{1}{2}}\left [\frac{\Gamma(\Delta_m-m-d+1)\Gamma(\Delta_m-m-\frac{d}{2})}{\Gamma(\Delta_m)\Gamma(\Delta_m -d+1)(2\Delta_m-d)}\right ]^{\frac{1}{2}}\nonumber\\
& & \times \left [\frac{\Gamma(\Delta-\frac{d}{2}+1+m)\Gamma(\Delta_1-\frac{d}{2}+1+m)\Gamma(\Delta+m)\Gamma(\Delta_1+m)}{\Gamma(m+1)\Gamma(\frac{d}{2}+m)}\right ]^{\frac{1}{2}}\nonumber\\
\end{eqnarray}
and we get the result for the anomalous dimension
\begin{eqnarray}
\delta \Delta_m &=&\frac{\lambda}{N^2}\frac{\Gamma(\frac{d}{2}+m)}{4\pi^{d/2}\Gamma(\frac{d}{2})\Gamma(m+1)}
\frac{\Gamma(\Delta_m-d+1)\Gamma(\Delta_m-m-\frac{d}{2})}{\Gamma(\Delta_m)\Gamma(\Delta_m-m-d+1)}\times\nonumber\\
& & \left [\frac{\Gamma(\Delta+m)\Gamma(\Delta_1+m)\Gamma(\Delta_2+m)\Gamma(\Delta_3+m)}{\Gamma(\Delta+m-\frac{d}{2}+1)\Gamma(\Delta_1+m-\frac{d}{2}+1)\Gamma(\Delta_2+m-\frac{d}{2}+1)\Gamma(\Delta_3+m-\frac{d}{2}+1)}\right ]^{1/2}\nonumber \\
\label{delm1}
\end{eqnarray}
This agrees (up to an overall coefficient) with results in \cite{Heemskerk:2009pn,Fitzpatrick:2011dm,Fitzpatrick:2010zm,Alday:2014tsa}. Here it seems the bulk bootstrap is more powerful since one does not need to know the conformal blocks.

One can also analyze other situations. For example suppose the first condition in (\ref{anomcond1}) is changed to $\Delta +\Delta_1=\Delta_2 +\Delta_3 +2k$. Then one can define for $0\leq m<k$
\begin{eqnarray}
({\cal O}_2{\cal O}_3)^{+}_{m}=\frac{1}{\sqrt{2}}[({\cal O}_2{\cal O}_3)_{m}]\nonumber\\
({\cal O}_2{\cal O}_3)^{-}_{m}=\frac{1}{\sqrt{2}}[({\cal O}_2{\cal O}_3)_{m}]
\end{eqnarray}
and for $m\geq k$
\begin{eqnarray}
({\cal O}_2{\cal O}_3)^{+}_{m}=\frac{1}{\sqrt{2}}[({\cal O}_2{\cal O}_3)_{m}+({\cal O}{\cal O}_1)_{m-k}]\nonumber\\
({\cal O}_2{\cal O}_3)^{-}_{m}=\frac{1}{\sqrt{2}}[({\cal O}_2{\cal O}_3)_{m}-({\cal O}{\cal O}_1)_{m-k}]
\label{anomks}
\end{eqnarray}
The bulk bootstrap still implies (\ref{bostpms}), but now this translates for $0\leq m<k$ into a condition on the ${\cal O}(1/N^2)$ OPE coefficients, while
for $m\geq k$ it gives a condition on the anomalous dimensions of the operators (\ref{anomks}).

\section{All scalar S-channel solutions\label{sect:alls}}
The method used in the previous section was useful for cases where one has ansatze for the bulk operator in different channels which are similar to each other and have a finite number of spins involved. However it becomes less useful when 
the ansatze in different channels are not similar, for instance when one exchanges a single-trace operator in one channel but not in other channels.

Suppose one has a situation where in one channel only a few spins contribute while in other channels many spins, or even all spins, are possible.
If one just wants to deduce the bulk operator, and not necessarily work out all the OPE coefficients or anomalous dimensions, it turns out there is a
simple way to proceed.
Let us start with the assumption that there is an OPE  channel in which only primary scalars contribute to the 4-point function (other channels can involve higher-spin primaries).
Let us assume this channel is the OPE between ${\cal O}_{2}(y_2)$ and ${\cal O}_{3}(y_3)$.
With this in mind we take the  OPE to be
\begin{equation}
{\cal O}_{2}(y_2){\cal O}_{3}(y_3)=d_{\chi}(y_2-y_3)\chi(y_2)+\sum_{m}d_{m23}(y_2-y_3)({\cal O}_{2}{\cal O}_{3})_{m}(y_2)+\sum_{n}c_{k}(y_2-y_3)({\cal O}{\cal O}_{1})_{k}(y_2)
\label{ope*}
\end{equation}
where we've assumed that only one single-trace operator $\chi(y)$ contributes.\footnote{Generalizations to more than one single-trace operator are straightforward.} In counting powers of $N$, note that $d_{\chi}\sim 1/N$, $d_{n23} \sim 1$ and $c_{k}\sim 1/N^2$.

We have an infinite sum of 3-point functions which we know how to deal with. For each primary operator appearing in\footnote{As before $({\cal O}{\cal O}_1)_k$ generically does not contribute causality-violating terms to this order.} 
\begin{equation}
d_{\chi}(y_2-y_3)\chi(y_2)+\sum_{m}d_{m23}(y_2-y_3)({\cal O}_{2}{\cal O}_{3})_{m}(y_2)
\end{equation}
we need to add a tower of higher-dimension primary scalar operators composed from the primary, ${\cal O}_{1}$ and derivatives, and add it to the definition of the bulk field. For instance starting from
 \begin{equation}
\langle{\cal O}(x) {\cal O}_{1}(y_1) \chi(y_2)\rangle
\end{equation}
with coefficient  $c_{\chi}$, we can correct the zeroth-order bulk operator $K{\cal O}(z,x)$ to respect locality by setting
\begin{equation}
\phi(z,x)=K{\cal O}(z,x)+\frac{1}{N}\sum_{n} a_{n} K_{n}({\cal O}_1 \chi)_{n}(z,x)
\end{equation}
where the coefficients $a_{n}$ are chosen to satisfy (\ref{ansol}). The corrected bulk field obeys the equation of motion
\begin{equation}
(\nabla^2-m_{0}^2)\phi(z,x)=\frac{\lambda_{\chi}}{N} K{\cal O}_{1}(z,x) K\chi (z,x)+\frac{1}{N}\sum^{l}_{k=0} \beta_{k}(\nabla^2-m_{0}^{2})^{k+1}K{\cal O}_{1}(z,x)K \chi (z,x)
\end{equation}
with  
\begin{equation}
\frac{\lambda_{\chi}}{N}=\frac{c_{\chi}}{\gamma_{F}(\Delta,\Delta_1\Delta_{\chi})}
\end{equation}
The coefficients $\beta_k$ are arbitrary and parametrize bulk field redefinitions.

Similarly we can start with a 3-point function
\begin{equation}
\langle{\cal O}(x) {\cal O}_{1}(y_1) ({\cal O}_{2}{\cal O}_{3})_{m}(y_2) \rangle  
\end{equation}
with coefficient $c_{m23}$.
To promote ${\cal O}(x)$ to a bulk operator which satisfies locality we must define the bulk field as 
\begin{equation}
\phi(z,x)=K{\cal O}(z,x)+\frac{1}{N^2}\sum_{n} a^{m}_{n} K_{n}({\cal O}_1 ({\cal O}_{2}{\cal O}_{3})_{m})_{n}(z,x)
\end{equation}
The coefficients $ a^{m}_{n}$ are chosen to obey (\ref{ansol})
with $M_{n}^2$ replaced by $M_{n,m}^2=\Delta_{n,m}(\Delta_{n,m}-d)$, where $\Delta_{n,m}=\Delta_{m}+\Delta_{1}+2n$ and
$\Delta_{m}=\Delta_{2}+\Delta_{3}+2m$.
The resulting local bulk field solves the equation of motion
\begin{eqnarray}
(\nabla^2-m_{0}^2)\phi(z,x)&=&\frac{\lambda_{m23}}{N^2}K{\cal O}_{1}(z,x) K({\cal O}_{2}{\cal O}_{3})_{m} (z,x)\nonumber\\
&+& \frac{1}{N^2}\sum^{l}_{k=0} \beta^{(m)}_{k}(\nabla^2-m_{0}^{2})^{k+1}K{\cal O}_{1}(z,x)K({\cal O}_{2}{\cal O}_{3})_{m}(z,x)\nonumber\\
\end{eqnarray}
with
\begin{equation}
\frac{\lambda_{m23}}{N^2}=\frac{c_{m23}}{{\gamma}_{F}(\Delta,\Delta_{1},\Delta_{m})}
\label{lamals}
\end{equation}
and $\beta_{k}^{(m)}$ arbitrary.

To see how to proceed it is easier to consider a concrete example.
Suppose we choose $\beta_{k}=0$ and $\beta_{k}^{(m)}=0$. This means we've defined a bulk operator, local in the $y_2 \rightarrow y_3$
OPE limit, so that it solves the equation of motion
\begin{equation}
(\nabla^2-m_{0}^2)\phi(z,x)=\frac{\lambda_{\chi}}{N}K{\cal O}_1(z,x)K \chi(z,x)+\frac{1}{N^2} K{\cal O}_1(z,x)\sum_{m}\lambda_{m23}K({\cal O}_{2}{\cal O}_{3})_{m}(z,x)
\label{eom1}
\end{equation}
Now that we have an ansatz for the bulk operator in the S-channel, we could demand that it be a local bulk operator in the full 4-point function (hence
automatically in all OPE channels). This however is a difficult condition to impose.  So instead imposing it on the bulk field directly, we impose it on $(\nabla^2-m^2)\phi(z,x)$: that is, on the combination appearing on the right-hand side of (\ref{eom1}). While this is not equivalent to demanding locality of the bulk
field, it is a necessary condition for the bulk field to be local.\footnote{Note that (\ref{eom1}) may not be the full equation of motion for $\phi$, 
if there are other non-trivial 4-point functions that contribute.}

So let us evaluate
\begin{equation}
\langle\left(\frac{\lambda_{\chi}}{N}K{\cal O}_1(z,x)K \chi(z,x)+\sum_{m}\frac{\lambda_{m23}}{N^2} K{\cal O}_1(z,x)K_{m} ({\cal O}_{2}{\cal O}_{3})_{m}(z,x)\right){\cal O}_{1}(y_1){\cal O}_{2}(y_2){\cal O}_{3}(y_3)\rangle 
\label{schan1}
\end{equation}
To leading order in $1/N$ this factorizes as
\be
\frac{1}{N}\langle K{\cal O}_1(z,x){\cal O}_{1}(y_1)\rangle \, \langle(\lambda_{\chi}K \chi(z,x)+\sum_{m}\frac{\lambda_{m23}}{N}K_{m} ({\cal O}_{2}{\cal O}_{3})_{m}(z,x)){\cal O}_{2}(y_2){\cal O}_{3}(y_3)\rangle 
\label{4point2}
\ee
The first factor in (\ref{4point2}) is a 2-point function which obeys causality.  In the second factor, we must distinguish between
$\lambda_{\chi}=0$ and $\lambda_{\chi} \neq 0$.

First we treat the case $\lambda_{\chi}=0$.  Then the second factor in (\ref{4point2}) is local if 
\begin{equation}
\lambda_{m23}=\sum^{k_{\rm max}}_{k=0} \rho_{k} M_{m}^{2k}b_{m23}
\label{scx0}
\end{equation}
with arbitrary coefficients $\rho_{k}$.  (We have introduced a cutoff $k_{\rm max}$ to avoid unbounded numbers of derivatives below.)
Assuming $\frac{1}{2}(\Delta+\Delta_1-\Delta_2-\Delta_3)$ is not an integer, from (\ref{lamals}), (\ref{scx0}), (\ref{coperel}) one gets the OPE
coefficients
\begin{equation}
\delta C^{{\cal O}{\cal O}_1}_{({\cal O}_2{\cal O}_3)_{m}}=\frac{1}{N^2}\frac{\pi^{\frac{d}{2}}\gamma_{F}(\Delta,\Delta_1,\Delta_m)\Gamma(\Delta_m-\frac{d}{2})}{(2\Delta_m -d)\Gamma(\Delta_m)}b_{m23}\sum^{k_{\rm max}}_{k=0} \rho_{k}(\Delta_m(\Delta_m-d))^{k}
\label{alsope}
\end{equation}
Taking $\rho_{0}=\lambda$ and $\rho_k=0$ for $k > 0$ we recover the results of the previous section. Again the relationship between the bootstrap and bulk locality is clear. Indeed (\ref{scx0}) corresponds to a local bulk interaction of the form
\begin{equation}
\sum_{k=0}^{k_{\rm max}} \rho_{k}  \phi \phi_{1}\nabla^{2k}\phi_{2}\phi_{3}(z,x).
\end{equation}
Alternatively, if $\frac{1}{2}(\Delta+\Delta_1-\Delta_2-\Delta_3)$ is an integer, then what gets fixed by the bootstrap are the anomalous dimensions, as in the previous section. For instance for $\Delta+\Delta_1=\Delta_2+\Delta_3$ one simply gets the result (\ref{delm1}) multiplied by $\sum_{k=0}^{k_{\rm max}} \rho_{k}(\Delta_m(\Delta_m-d))^{k}$.

On the other hand if $\lambda_{\chi} \neq 0$ then the right-hand side of (\ref{4point2}) has the form of a bulk-boundary 3-point function,
where the bulk field is
\begin{equation}
\lambda_{\chi}K \chi(z,x)+\frac{1}{N}\sum_{m}\lambda_{m23}K_{m} ({\cal O}_{2}{\cal O}_{3})_{m}(z,x)
\end{equation}
We know from (\ref{ansol}) that for this to be local one needs to have (taking for simplicity $\beta_k=0$)
\begin{equation}
\frac{1}{N}\frac{\lambda_{m23}}{\lambda_{\chi}}= \alpha_{\chi}\frac{b_{m23}}{M_{m}^2-m_{\chi}^{2}}
\label{scxn0}
\end{equation}
with $M_{m}^{2}=(\Delta_2 +\Delta_3 +2m)(\Delta_2 +\Delta_3 +2m-d)$. The coefficient $\alpha_{\chi}$ is
\begin{equation}
\alpha_{\chi}=\frac{c_{\chi}^{(23)}}{\gamma_{F}(\Delta_{\chi},\Delta_2,\Delta_3)}
\end{equation}
where $c_{\chi}^{(23)}$ is the coefficient appearing in the 3-point function $\langle\chi(x){\cal O}_{2}(y_2){\cal O}_{3}(y_3)\rangle $.
The OPE coefficient\footnote{Alternatively this procedure could determine anomalous dimensions, if the dimensions of the operators involved obey
relations such as (\ref{anomcond1}).} for this case is now
\begin{equation}
\delta C^{{\cal O}{\cal O}_1}_{({\cal O}_2{\cal O}_3)_{m}}=\frac{1}{N}\frac{\lambda_{\chi}\alpha_{\chi}}{M_{m}^{2}-m_{\chi}^{2}}\frac{\pi^{\frac{d}{2}}\gamma_{F}(\Delta,\Delta_1,\Delta_m)\Gamma(\Delta_m-\frac{d}{2})}{(2\Delta_m -d)\Gamma(\Delta_m)}b_{m23}
\label{alsope1}
\end{equation}
Of course one can add to this result the terms in (\ref{alsope}).
We see that in the simple case with no extra contact interaction
\begin{equation}
\delta C^{{\cal O}{\cal O}_1}_{({\cal O}_2{\cal O}_3)_{m}}(c_{\chi}\neq 0)\sim \frac{1}{M_{m}^{2}-m_{\chi}^{2}}\delta C^{{\cal O}{\cal O}_1}_{({\cal O}_2{\cal O}_3)_{m}}(c_{\chi}=0)
\end{equation}
This relationship was recently noted in \cite{Hijano:2015zsa}, from computations of bulk geodesic Witten diagrams representing conformal blocks.

The solution (\ref{scxn0}) corresponds to a bulk cubic coupling
\begin{equation}
\phi \phi_{1}\chi(z,x)+\phi _{2}\phi_{3}\chi(z,x)
\end{equation}
To see this consider the equations of motion for the fields $\phi(x,z)$ and $\chi(x,z)$.
\begin{eqnarray}
(\nabla^2-m^2)\phi(x,z)&=&\frac{1}{N}\phi_{1}\chi(x,z)\nonumber\\
(\nabla^2-m_{\chi}^2)\chi(x,z)&=&\frac{1}{N}(\phi_{1}\phi(x,z)+\phi_{2}\phi_{3}(x,z))
\end{eqnarray}
Let us now expand the bulk fields as
\begin{equation}
\phi_{j}(x,z)=\phi_{j}^{(0)}+\frac{1}{N}\phi_{j}^{(1)}+ \cdots
\end{equation}
where $\phi_{j}^{(0)}=K{\cal O}_{j}(x,z)$. 
To ${\cal O}({1}/{N})$ one has
\begin{eqnarray}
(\nabla^2-m^2)\phi^{(1)}(x,z)&=&\phi_{1}^{(0)}\chi^{(0)}(x,z)\label{phi1eom}\\
(\nabla^2-m_{\chi}^2)\chi^{(1)}(x,z)&=&\phi^{(0)}_{1}\phi^{(0)}(x,z)+\phi^{(0)}_{2}\phi^{(0)}_{3}(x,z)\nonumber
\end{eqnarray}
These equations are solved by
\begin{eqnarray}
&&\phi^{(1)}(z,x)=\sum_{n}\frac{b_{n1\chi}}{M_{n}^2-m^2}K({\cal O}_{1}\chi)_{n}(z,x)\nonumber\\
&&M_{n}^{2}=\Delta_n(\Delta_n -d),\ \ \ \Delta_{n}=\Delta_1+\Delta_{\chi}+2n
\end{eqnarray}
and
\begin{eqnarray}
&&\chi^{(1)}(z,x)=\sum_{n}\frac{b_{n1\phi}}{(M^{(1)}_{n})^2-m_{\chi}^2}K({\cal O}_{1}{\cal O})_{n}(z,x) +\sum_{n}\frac{b_{n23}}{(M^{(23)}_{n})^2-m_{\chi}^2}K({\cal O}_{2}{\cal O}_3)_{n}(z,x)\nonumber\\
&&(M_{n}^{(1)})^2 = \Delta^{(1)}_{n}(\Delta^{(1)}_{n} -d) ,  \ \   \Delta^{(1)}_{n}=\Delta_1+\Delta+2n \label{chieom}
\end{eqnarray}
Then to ${\cal O}(1/N^2)$ one gets
\begin{equation}
(\nabla^2-m^2)\phi^{(2)}(x,z)=\phi_{1}^{(0)}\chi^{(1)}(x,z)+\phi_{1}^{(1)}\chi^{(0)}(x,z)
\label{eomphi2}
\end{equation}
The 4-point function and solution in (\ref{eom1}) and (\ref{scxn0}) corresponds to the second term on the right of (\ref{chieom})
inserted in the first term on the right of (\ref{eomphi2}), plus of course the ${\cal O}({1}/{N})$ term (\ref{phi1eom}).
Other solutions for $\lambda_{m23}$ are possible. We know from \cite{Kabat:2015swa} that multiplying $\lambda_{m23}$ in (\ref{scxn0}) by $({(M^{(23)}_{m})^{2}}/{m_{\chi}^{2}})^{k}$ also works, and corresponds to bulk interactions 
\begin{equation}
\phi \phi_{1}\chi(z,x)+ \chi(z,x)(\frac{\nabla^2}{m_{\chi}^{2}})^{k}\phi _{2}\phi_{3}(z,x)
\end{equation}

\section{Vector exchange\label{sect:allv}}
In this section we repeat the previous analysis of scalar exchange for the case where the exchanged operator is a primary non-conserved current
with spin one.

\subsection{Preliminaries\label{sect:prev}}
We start with a 3-point function involving a scalar of dimension $\Delta$, another scalar of dimension $\Delta_{j}$, and a primary current $j_{\mu}$ with
dimension $\Delta_{\rm v}>d-1$.
\begin{eqnarray}
\label{3pointv}
&&\langle{\cal O}(x){\cal O}_{j}(y_1)j_{\mu}(y_2)\rangle =\left ( \frac{(y_1-y_2)_{\mu}}{(y_1-y_2)^2}- \frac{(x-y_2)_{\mu}}{(x-y_2)^2}\right ) \\
\nonumber
&&\qquad \times \frac{\gamma^{({\rm v})}}{(y_1-x)^{\Delta+\Delta_j-\Delta_{\rm v}+1}(y_1-y_2)^{\Delta_j+\Delta_{\rm v} -\Delta-1}(x-y_2)^{\Delta-\Delta_j+\Delta_{\rm v}-1}}
\end{eqnarray}
We want to promote ${\cal O}(x)$ to a  local bulk field so we start with 
\begin{equation}
\langle K{\cal O}(z,x) {\cal O}_{j}(y_1)j_{\mu}(y_2)\rangle 
\end{equation}
This 3-point function does not obey bulk locality, but
we can make it local by adding a tower of appropriately-smeared higher-dimension primary scalar operators, built from ${\cal O}_{j}$, $j_{\mu}$ and derivatives, to the definition of the bulk field.  The coefficients of these operators are fixed by requiring bulk causality.
Then we can compute  $(\nabla^2-m_{0}^2)\phi(z,x)$ and find (see appendix B and \cite{Kabat:2015swa})
\begin{equation}
(\nabla^2-m_{0}^2)\phi(z,x)=-\frac{\gamma^{({\rm v})}}{\gamma_{F}(\Delta,\Delta_{j}+1,\Delta_{\rm v})}\frac{\Delta_{\rm v}}{(\Delta_j-\frac{d}{2})(\Delta_{\rm v}+\Delta-\Delta_j-1)}\partial^{M}K{\cal O}_{j}(z,x)A^{(0)}_{M}(z,x)
\label{basicveceom}
\end{equation}
where $M=(z,x)$ and 
$A^{(0)}_{M}(z,x)$ is a bulk massive vector field given in terms of CFT operators by \cite{Kabat:2012hp}
\begin{eqnarray}
\label{mvecsmear}
&&A^{(0)}_{z}(z,x)=\frac{1}{d-\Delta_{\rm v}-1}K_{\Delta_{\rm v}}(\partial^{\mu}j_{\mu})(z,x) \\
&&zA^{(0)}_{\mu}(z,x)=K_{\Delta_{\rm v}}j_{\mu}(z,x) +\frac{z}{2(\Delta_{\rm v}-\frac{d}{2}+1)}\frac{1}{d-\Delta_{\rm v}-1}K_{\Delta_{\rm v}+1}\partial_{\mu}(\partial^{\nu}j_{\nu})(z,x)\nonumber
\end{eqnarray}
Here we have indicated by $K_{\Delta}$ the smearing appropriate to a scalar of dimension $\Delta$.

We will need another result, developed in appendix C.1.  Start with two single-trace scalar primaries ${\cal O}_{j}$ and ${\cal O}_k$ of equal dimension,
$\Delta_j=\Delta_k$. We can build to leading order in $1/N$ an infinite tower of double-trace non-conserved primary currents out of
${\cal O}_{j}$, ${\cal O}_k$ and derivatives. We label the resulting current of dimension $\Delta_{m}=\Delta_j +\Delta_k +1+2m$ by $({\cal O}_{j}{\cal O}_{k})_{\mu}^{m}$.
We can smear these currents as in (\ref{mvecsmear}) to obtain bulk operators whose 2-point functions coincide with the 2-point function of a free  massive vector field. We label this bulk operator by $K({\cal O}_{j}{\cal O}_{k})_{M}^{m}(z,x)$, where $M=(z,x)$ is a bulk spacetime vector index.
Then there is a set of coefficients $\tilde{b}_{mjk}^{(\rm v)}$ such that
\begin{equation}
(2\Delta_j-d) \sum \tilde{b}^{({\rm v})}_{mjk}K({\cal O}_{j}{\cal O}_{k})^{m}_{M}(z,x)=\phi^{(0)}_{k}(z,x)\partial_{M}\phi^{(0)}_{j}(z,x)-\phi^{(0)}_{j}(z,x)\partial_{M}\phi^{(0)}_{k}(z,x)
\label{basicbv}
\end{equation}

\subsection{Vector example\label{sect:4pointv}}
Here we solve the bootstrap in an example where in one channel non-conserved primary double-trace currents can be exchanged.
We assume for simplicity that in a different channel $y_2 \rightarrow y_3$ only scalar double-trace operators contribute. This means
the equation of motion for the bulk field has the general form
\begin{eqnarray}
\label{opev1}
(\nabla^2-m_{0}^2)\phi(z,x) &=& \frac{1}{N^2}K{\cal O}_{1}(z,x) \sum_{m}\lambda_{m}^{(1)}K({\cal O}_{2}{\cal O}_{3})_{m}(z,x) \\
&+&\frac{1}{N^2} \sum_{k=1}^{l_1}(\nabla^2-m_{0}^2)^{k}\left ( K{\cal O}_{1}(z,x)\sum_{m}\beta^{(1)}_{(k)m}K({\cal O}_{2}{\cal O}_{3})_{m}(z,x)\right ) \nonumber
\end{eqnarray}
Here $\beta_{(k)m}^{(1)}$ parametrize field redefinitions and are not determined by CFT data,
while $\lambda_{m}^{(1)}$ is related to CFT data as in (\ref{lamcerel}).
In the OPE channel $y_1 \rightarrow y_3$ we assume that both scalar and vector double-trace operators contribute.
Vector double-trace operators have a 3-point function
\begin{eqnarray}
\label{3pointvssn}
&&\langle{\cal O}(x){\cal O}_{2}(y_1)({\cal O}_1{\cal O}_{3})^{m}_{\mu}(y_2)\rangle =\left ( \frac{(x-y_2)_{\mu}}{(x-y_2)^2}- \frac{(y_1-y_2)_{\mu}}{(y_1-y_2)^2}\right ) \\
\nonumber
&& \qquad \times \frac{\gamma^{({\rm v})}_{m}}{(x-y_1)^{\Delta + \Delta_2-\Delta_m+1}(y_1-y_2)^{\Delta_2 - \Delta + \Delta_m-1}(x-y_2)^{\Delta - \Delta_2 + \Delta_m-1}}
\end{eqnarray}
where $\Delta_m=\Delta_1 +\Delta_3+2m+1$. 
Thus in this channel the equation of motion for the bulk field is (see section \ref{sect:prev})
\begin{eqnarray}
(\nabla^2-m_{0}^2)\phi(z,x) &=& \frac{1}{N^2}K{\cal O}_{2}(z,x) \sum_{m}\lambda_{m}^{(2)}K({\cal O}_{1}{\cal O}_{3})_{m}(z,x) \nonumber\\
&+& \frac{1}{N^2}\sum_{k=1}^{l_2}(\nabla^2-m_{0}^2)^{k}\left ( K{\cal O}_{2}(z,x)\sum_{m}\beta^{(2)}_{(k)m}K({\cal O}_{1}{\cal O}_{3})_{m}(z,x)\right )\nonumber\\
&+& \frac{1}{N^2}\partial^{M}K{\cal O}_{2}(z,x)\sum_{m}\lambda^{({\rm v})}_{m} A^{m}_{M}(z,x)
\label{opev2}
\end{eqnarray}
Here $A_{M}^{m}(z,x)$ is the appropriate smearing of the non-conserved primary current
$({\cal O}_{1}{\cal O}_{3})^{m}_{\mu}$ of dimension $\Delta_m=\Delta_1+\Delta_3+2m+1$ to make a bulk free massive vector. Again $\beta_{(k)m}^{(2)}$ parametrizes field redefinitions and is not determined by CFT data, while $\lambda_{m}^{(2)}$ is given by (see (\ref{lamcerel}))
\begin{equation}
\frac{\lambda_{m}^{(2)}}{N^2}=\frac{c_{m13}}{\gamma_{F}(\Delta,\Delta_2,\Delta_1+\Delta_3 +2m)}
\end{equation}
and $\lambda_{m}^{({\rm v})}$ is given by (see (\ref{basicveceom}))
\begin{equation}
\frac{\lambda_{m}^{({\rm v})}}{N^2}= -\frac{\gamma_{m}^{({\rm v})}}{\gamma_{F}(\Delta,\Delta_{2}+1,\Delta_{m})}\frac{\Delta_m}{(\Delta_2-\frac{d}{2})(\Delta-\Delta_2+\Delta_{m}-1)}.
\label{lamv}
\end{equation}
If we assume a similar expansion in the $y_1 \rightarrow y_2$ channel we will get the same result with $2 \leftrightarrow 3$.

Comparing (\ref{opev1}) and (\ref{opev2}), we see that for them to agree (which they must according to the bootstrap) all the sums over $m$ must give local bulk expressions in terms of $K{\cal O}_{i}(z,x)$. As a simple example we limit ourselves to expressions that have at most two bulk derivatives on the right-hand side of the equation of motion for $\phi$, and we take $\beta_{(k) m}^{(1)}=0$. A solution to the bulk bootstrap is then parametrized by
(we assume $\Delta_1=\Delta_3$)
\begin{eqnarray}
&&\sum_{m}\lambda_{m}^{(1)}K({\cal O}_{2}{\cal O}_{3})_{m}(z,x)=\lambda_1K{\cal O}_{2}(z,x)K{\cal O}_{3}(z,x)+\lambda_2\nabla^2 K{\cal O}_{2}(z,x)K{\cal O}_{3}(z,x)\nonumber\\
&&\sum_{m}\lambda_{m}^{(2)}K({\cal O}_{1}{\cal O}_{3})_{m}(z,x)=\alpha_1 K{\cal O}_{1}(z,x)K{\cal O}_{3}(z,x)+\alpha_2 \nabla^2 K{\cal O}_{1}(z,x)K{\cal O}_{3}(z,x)\nonumber\\
&&\sum_{m}\beta^{(2)}_{(1)m}K({\cal O}_{1}{\cal O}_{3})_{m}(z,x)=\beta K{\cal O}_{1}(z,x)K{\cal O}_{3}(z,x)\\
&&\sum_{m}\lambda^{({\rm v})}_{m} A^{m}_{M}(z,x)=\gamma (K{\cal O}_{1}(z,x) \partial_{M} K{\cal O}_{3}(z,x)-K{\cal O}_{3}(z,x) \partial_{M} K{\cal O}_{1}(z,x))\nonumber
\end{eqnarray}
This means that
\begin{eqnarray}
&&\lambda_{m}^{(1)}=\lambda_1 b_{m23} +\lambda_2 (M^{(23)}_{m})^{2} b_{m23}\nonumber\\
&&\lambda_{m}^{(2)}=\alpha_1 b_{m13} +\alpha_2 (M^{(13)}_{m})^{2} b_{m13}\\
&&\beta^{(2)}_{(1)m}=\beta b_{m13}\nonumber\\
&&\lambda^{({\rm v})}_{m}=\gamma(2\Delta_1 -d)\tilde{b}^{({\rm v})}_{m31}\nonumber
\end{eqnarray}
where  $(M^{(ij)}_{m})^{2}=(\Delta_i +\Delta_j +2m)(\Delta_i +\Delta_j +2m -d)$ and $\tilde{b}^{({\rm v})}_{m31}$ is described in (\ref{basicbv}).
After a little algebra one finds that the two expressions for the equation of motion agree if
\begin{eqnarray}
&&\alpha_{1}=\lambda_2(m_{3}^{2}+\frac{m_{0}^2+m_{2}^{2}}{2})+\lambda_1 \nonumber\\
&&-\alpha_{2}=\beta=\frac{\gamma}{2}=\frac{\lambda_2}{2}
\end{eqnarray}
which implies a bulk quartic interaction
\begin{equation}
\lambda_1\phi\phi_1 \phi_2 \phi_3(z,x)+\lambda_2\phi\phi_1\nabla^2 ( \phi_2 \phi_3)(z,x)
\end{equation}

This solution for the bootstrap computes a variety of OPE coefficients to order $1/N^2$, since all $\lambda_{m}^{(*)}$ are linearly related to some OPE coefficient. In fact $\lambda_{m}^{(1)}$ and $\lambda_{m}^{(2)}$ are related to the OPE coefficients for two scalars going to a double-trace primary scalar, as shown
in (\ref{coperel}), (\ref{lamcerel}). But $\lambda_{m}^{({\rm v})}$  on the other hand is a new entity. To see its connection to an OPE coefficient  we look at the OPE of ${\cal O}(x)$ and ${\cal O}_{2}(y_2)$ producing primary currents.
\begin{equation}
{\cal O}(x){\cal O}_{2}(y_2)=\cdots +\sum_{m}\delta C^{{\cal O}{\cal O}_{2}}_{({\cal O}_1{\cal O}_{3})^{m}_{\mu}}\frac{(y_2-x)^{\nu}}{(x-y_2)^{\Delta + \Delta_2+1-\Delta_{m}}}({\cal O}_1{\cal O}_{3})^{m}_{\nu}(x)+\cdots
\end{equation}
Comparing this to (\ref{3pointvssn}) and with the normalization used in (\ref{lamv})
\begin{equation}
\langle({\cal O}_1{\cal O}_{3})^{m}_{\mu}(x)({\cal O}_1{\cal O}_{3})^{m}_{\nu}(y_2)\rangle =\frac{(2\Delta_m -d)\Gamma(\Delta_{m})}{\pi^{d/2}\Gamma(\Delta_m -\frac{d}{2})}\frac{\left ( \eta_{\mu\nu}-2\frac{(x-y_2)_{\mu}(x-y_2)_{\nu}}{(x-y_2)^2}\right)
}{(x-y_2)^{2\Delta_m}}
\end{equation}
we find that the OPE coefficient is just
\begin{equation}
\delta C^{{\cal O}{\cal O}_{2}}_{({\cal O}_1{\cal O}_{3})^{m}_{\mu}}=-\gamma^{({\rm v})}_{m}\frac{\pi^{d/2}\Gamma(\Delta_m -\frac{d}{2})}{(2\Delta_m-d)\Gamma(\Delta_{m})}
\label{delcv}
\end{equation}
which gives
\begin{equation}
\delta C^{{\cal O}{\cal O}_{2}}_{({\cal O}_1{\cal O}_{3})^{m}_{\mu}}=\frac{\gamma}{N^2}\frac{\pi^{d/2}(\Delta_2-\frac{d}{2})\Gamma(\Delta_m -\frac{d}{2})}{(2\Delta_m-d)\Gamma(\Delta_{m}+1)}\gamma_{F}(\Delta,\Delta_2+1,\Delta_m)(\Delta_m+\Delta-\Delta_2-1)(2\Delta_1-d)\tilde{b}^{({\rm v})}_{m31}
\end{equation}

\subsection{All spin-$1$ S-channel solutions\label{sect:allv1}}
In this section we analyze a scalar 4-point function where in one channel the exchange is limited to spin one, either single-trace or double-trace,
but in other channels the spins are unbounded.  We start with the 4-point function
\begin{equation}
\langle K{\cal O}(z,x){\cal O}_1(y_1){\cal O}_2(y_2){\cal O}_3(y_3)\rangle 
\end{equation}
We assume that the OPE of ${\cal O}_2(y_2)$ with ${\cal O}_3(y_3)$ has the form
\begin{equation}
{\cal O}_2(y_2){\cal O}_3(y_3)=\beta^{\mu}(y_2-y_3)\chi_{\mu}(y_2)+ \sum_{m}\alpha_{m}^{\mu}(y_2-y_3)({\cal O}_2{\cal O}_3)_{\mu}^{m}(y_2) +\cdots
\label{vecope}
\end{equation}
where $\chi_{\mu}$ is the only single-trace primary non-conserved current present in the OPE, and 
$({\cal O}_2{\cal O}_3)_{\mu}^{m}$ is a tower of double-trace non-conserved primary currents.

The OPE reduces the original 4-point function to an infinite sum of 3-point functions involving two scalars and a non-conserved primary current.
We know from section \ref{sect:prev} that for each primary current
we can add an infinite tower of smeared triple-trace primary scalars to $K{\cal O}(z,x)$, constructed from ${\cal O}_1$, the primary current and derivatives, to make
each 3-point function local.  The resulting bulk operator, which appears local in this OPE limit, obeys the equation of motion (see (\ref{basicveceom}))
\begin{equation}
(\nabla^2-m^2)\phi(z,x)=\frac{\lambda^{({\rm v})}_{\chi}}{N}\partial^{M}K{\cal O}_{1}(z,x)A^{(\chi)}_{M}(z,x)+\frac{1}{N^2}\sum_{m} \lambda_{m23}^{({\rm v})}\partial^{M}K{\cal O}_{1}(z,x)A^{(m)}_{M}(z,x)_{m}
\label{beomv}
\end{equation}
Here
$A^{(\chi)}_{M}(z,x)$ is the bulk massive vector one gets by smearing the single-trace primary current $\chi_{\mu}$ as in (\ref{mvecsmear}).  Likewise
$A_{M}^{(m)}(z,x)$ is the bulk massive vector one gets by smearing the double-trace primary current $({\cal O}_2{\cal O}_3)_{\mu}^{m}$ as in (\ref{mvecsmear}).
Finally $\lambda^{({\rm v})}_{\chi}$ and $\lambda^{({\rm v})}_{m23}$ are given by 
\begin{eqnarray}
&&\frac{\lambda^{({\rm v})}_{\chi}}{N}= -\frac{c^{({\rm v})}_{\chi}}{\gamma_{F}(\Delta,\Delta_{1}+1,\Delta_{\chi})}\frac{\Delta_{\chi}}{(\Delta_1-\frac{d}{2})(\Delta-\Delta_1+\Delta_{\chi}-1)}\nonumber\\
&&\frac{\lambda^{({\rm v})}_{m23}}{N^2}= -\frac{c^{({\rm v})}_{m23}}{\gamma_{F}(\Delta,\Delta_{1}+1,\Delta_{m})}\frac{\Delta_{m}}{(\Delta_1-\frac{d}{2})(\Delta-\Delta_1+\Delta_{m}-1)}
\label{eomcoef}
\end{eqnarray}
Here $c^{({\rm v})}_{\chi}$ and $c^{({\rm v})}_{m23}$ are the coefficients of the 3-point function (as in (\ref{3pointv})) of ${\cal O}(x)$, ${\cal O}_{1}(y_1)$ and either $\chi_{\mu}(y_2)$ or $({\cal O}_2{\cal O}_3)_{\mu}^{m}(y_2)$.
However this construction gives a local bulk field in a particular OPE limit $y_2 \rightarrow y_3$.  We are not guaranteed that the field will be local away from this
limit, or that the sum over the infinite series of 3-point functions will converge.

One way to check that everything works is to insert the expression for $\phi(z,x)$ into the full 4-point function and see what we get.  A simpler but necessary
condition is to impose bulk locality, not on $\phi$ itself, but on $(\nabla^2-m^2)\phi(z,x)$.  So we will require that the 4-point function
\begin{equation}
\langle(\nabla^2-m^2)\phi(z,x){\cal O}_{1}(y_1){\cal O}_{2}(y_2){\cal O}_{3}(y_3)\rangle 
\end{equation}
obeys bulk causality.
To leading order in $1/N$ and using (\ref{beomv}) we find that this 4-point function factorizes as
\begin{equation}
\langle\partial^{M}K{\cal O}_{1}(z,x){\cal O}_{1}(y_1)\rangle \, \langle(\frac{\lambda^{({\rm v})}_{\chi}}{N}A^{(\chi)}_{M}(z,x)+\sum_{m}\frac{\lambda_{m23}^{({\rm v})}}{N^2}A^{(m)}_{M}(z,x)_{m}) {\cal O}_{2}(y_2){\cal O}_{3}(y_3)\rangle 
\label{sv4point1}
\end{equation}
The first factor is a 2-point function that obeys bulk causality, so we only need to worry about the second factor.
If $\lambda_{\chi}^{({\rm v})}=0$ then the condition that the second factor in (\ref{sv4point1}) obeys bulk causality has a simple solution,
\begin{equation}
\lambda_{m23}^{({\rm v})}=\lambda b_{m23}^{({\rm v})}
\end{equation}
where $b_{m23}^{({\rm v})}$ are the coefficients defined in appendix C.
If $\lambda_{\chi}^{({\rm v})} \neq 0$, the condition for the second factor to obey causality is just that 
\begin{equation}
\lambda^{({\rm v})}_{\chi}A^{(\chi)}_{M}(z,x)+\frac{1}{N}\sum_{m} \lambda_{m23}^{({\rm v})}A^{(m)}_{M}(z,x)_{m}
\end{equation}
behaves as a local bulk massive vector field inside a 3-point function.
The condition for this is analyzed in appendix C. If we label the coefficient in the 3-point function $\langle\chi_{\mu} {\cal O}_{2} {\cal O}_{3}\rangle $ by  $d_{\chi}^{({\rm v})}$, then the result for $\Delta_2=\Delta_3$ is that 
\begin{equation}
\frac{\lambda^{({\rm v})}_{m23}}{N}= d_{\chi}^{({\rm v})}\lambda_{\chi}^{({\rm v})}\frac{b_{m23}^{({\rm v})}}{\Delta_{m}(\Delta_{m}-d)-\Delta_{\chi}(\Delta_{\chi}-d)}
\end{equation}
produces a local bulk massive vector field.

Finally let's see how the coefficients $\lambda_{m23}^{({\rm v})}$ are related to OPE coefficients in the CFT.
As in (\ref{delcv}) the OPE coefficient is just
\begin{equation}
\delta C^{{\cal O}{\cal O}_{1}}_{({\cal O}_2{\cal O}_{3})^{m}_{\mu}}=-c^{({\rm v})}_{m23}\frac{\pi^{d/2}\Gamma(\Delta_m -\frac{d}{2})}{(2\Delta_m -d)\Gamma(\Delta_{m})}
\end{equation}
We can now find an expression for the OPE coefficient. If $\lambda_{\chi}^{\rm v}=0$ then
\begin{equation}
\delta C^{{\cal O}{\cal O}_{1}}_{({\cal O}_2{\cal O}_{3})^{m}_{\mu}}=\frac{\pi^{d/2}\Gamma(\Delta_m -\frac{d}{2})}{(2\Delta_m -d)\Gamma(\Delta_{m})}\frac{(\Delta_1-\frac{d}{2})(\Delta_{m}+1)}{\Delta_m}\gamma_{F}(\Delta_1,\Delta_1 +1,\Delta_m)\frac{\lambda b_{m23}^{({\rm v})}}{N^2}
\label{vope}
\end{equation}
while if $\lambda_{\chi}^{\rm v}\neq 0$ then in (\ref{vope}) we need to replace
\begin{equation}
\frac{\lambda b_{m23}^{({\rm v})}}{N^2} \rightarrow \frac{d_{\chi}^{({\rm v})}\lambda_{\chi}^{({\rm v})}}{N}\frac{b_{m23}^{({\rm v})}}{\Delta_m(\Delta_m-d)-\Delta_{\chi}(\Delta_{\chi}-d)}
\end{equation}
Again the bootstrap gave us quite easily the OPE coefficients to ${\cal O}(1/N^2)$.  We also see very explicitly how these OPE coefficients build up the bulk equations of motion.

\section{Conclusions}
In this paper we further developed the construction of local bulk operators in the CFT. We extended our previous work on 3-point functions and
constructed bulk operators  which obey locality inside a 4-point function. This was done by using the OPE to reduce the 4-point function to an infinite sum
of 3-point functions.  Our previous results on 3-point functions \cite{Kabat:2015swa} guarantee that in each OPE channel one can define a local bulk operator. However consistency conditions, stemming from the overlap of OPE expansions in different channels, require that all these seemingly-different bulk operators actually agree up to bulk field redefinitions. These consistency conditions, which we refer to as the bulk bootstrap, turn out to determine either OPE coefficients or anomalous dimensions of double-trace operators, in agreement with results in the literature. Thus the bulk bootstrap not only determines the CFT expression for a bulk operator, it also provides a way to fix CFT data.  In many situations the bulk bootstrap is simpler
to solve than the regular conformal bootstrap, in part because to satisfy bulk locality we only need to consider primary fields, which means expressions for conformal blocks are not required.
In situations where only double-trace operators are exchanged, our analysis makes it clear that the bootstrap conditions can only be solved by bulk
fields which obey local equations of motion.  Thus the bootstrap implies that the CFT has a local gravity dual \cite{Heemskerk:2009pn}. Our analysis also simply and directly connects CFT data (OPE coefficients or anomalous dimensions of double-trace operators) to bulk equations of motion.

We also analyzed a class of examples where, in addition to double-trace primaries, a single-trace operator (either scalar or vector) can
contribute in one channel of the OPE.
Since in this case one expects that in other channels double-trace operators of all spins will contribute, it was more efficient to consider the bulk operator  constructed in this channel only, and to demand that it satisfy bulk locality away from this particular OPE limit.
This procedure enabled us to determine the bulk operator, and to read off the OPE coefficients of the double-trace operators
in this channel, in a much wider class of examples.

Note that in all cases the bulk bootstrap provides a simple method to determine the ${\cal O}(1/N^2)$ corrections to either OPE coefficients or
anomalous dimensions. What we mean by this is that the equation of motion (or the bulk interaction) directly encodes these quantities. For generic
operator dimensions the bootstrap determines OPE coefficients, and to ${\cal O}(1/N^2)$ there are no anomalous dimensions. But when
operator dimensions differ by integers the bootstrap turns out to determine anomalous dimensions, as described in section \ref{sect:4pointan}.
In this case there are also corrections to OPE coefficients at ${\cal O}(1/N^2)$, however these corrections are not encoded in a simple way in
the bulk interaction. This does not mean that they are not determined by the bootstrap: the anomalous dimensions already determine the bulk
interaction, so one could use this knowledge to compute the boundary 4-point function, and from this deduce the corrections to OPE coefficients.
This is in line with results in \cite{Heemskerk:2009pn}, where corrections to OPE coefficients could be computed from results for anomalous dimensions.

In this paper we studied examples involving exchanges of scalars and non-conserved currents. As directions for future work clearly generalizing to higher spin exchange is important, but also note that we did not treat exchange of a conserved current which brings in additional subtleties.
The construction of bulk operators was carried out in cases where the dual bulk diagrams are tree level 4-point functions. From the construction it seems clear that treating higher-point functions at tree level is straightforward. A more challenging question is how to treat effects that result
from bulk loops. We hope to comment on this in the near future.  Finally it is generally believed that a local bulk theory can emerge only if there is a large gap in the spectrum of the single-trace operator dimensions \cite{Heemskerk:2009pn}. It would be interesting to understand where this requirement
appears in the approach followed here.

\bigskip\bigskip
\goodbreak
\centerline{\bf Acknowledgements}
\noindent
We thank the KITP for hospitality during the program ``Quantum Gravity Foundations: UV to IR'' where this work was initiated.  At KITP this work was supported in part by NSF grant PHY11-25915.  We also thank the City College physics department for hospitality during this work.
DK is supported by U.S.\ National Science Foundation grant PHY-1519705 and is grateful to the Columbia University Center for Theoretical Physics for
hospitality.  The work of GL is supported in part by the Israel Science
Foundation under grant  504/13 and in part by a grant from
GIF, the German-Israeli Foundation for Scientific Research and
Development, under grant 1156-124.7/2011.

\appendix
\section{Scalar 3-point function}
We want to promote ${\cal O}(x)$ to a bulk operator inside a 3-point function, so we start with
\begin{equation}
\langle\phi^{(0)}(z,x) {\cal O}_{j}(y_1) {\cal O}_{k}(y_2)\rangle =\frac{1}{(y_1 - y_2)^{2\Delta_j}}
\left[\frac{z}{z^2+(x-y_2)^2}\right]^{\Delta_k-\Delta_j} I(\chi)
\label{3point1}
\end{equation}
Here 
\be
\label{chidef2}
\chi=\frac{[(x-y_1)^2+z^2][(x-y_2)^2+z^2]}{(y_1-y_2)^2 z^2}
\ee
and
\begin{equation}
\label{Idef2}
I(\chi) = \frac{\tilde{\gamma}}{2\Delta-d} \left(\frac{1}{\chi-1}\right)^\DS F\big(\,\DS,\,\DS - \frac{d}{2} + 1,\, \Delta_i - \frac{d}{2} + 1,\,\frac{1}{1-\chi}\,\big)
\end{equation}
with $\Delta_{\textstyle *}=\frac{1}{2}(\Delta+\Delta_{j}-\Delta_{k})$.  If $\DS$ or $\Delta - \DS$ are either zero or a negative integer then the hypergeometric series terminates
and (\ref{Idef2}) is analytic about $\chi = 1$.  But otherwise this expression does not obey bulk causality, due to the cut at $0 < \chi < 1$ which corresponds to bulk spacelike separation.
The discontinuity across the cut is given by (dropping a factor of $2\pi i$)
\begin{equation}
I_{\rm discont}=\frac{\frac{\tilde{\gamma}}{2\Delta-d}}{(1-\chi)^{\frac{d}{2}-1}} \frac{\Gamma(\Delta-\frac{d}{2}+1)}{\Gamma(\DS)\Gamma(\Delta-\DS)\Gamma(2-\frac{d}{2})} F\big(\,\DS-\frac{d}{2}+1,\,1+\DS-\Delta,\,2-\frac{d}{2},\,1-\chi\,\big)
\label{discont}
\end{equation}

To restore bulk causality we need to change the definition of the bulk operator.
The change is given by adding to the zeroth order definition (\ref{zerodef}) a tower of higher-dimension primary double-trace operators built from
${\cal O}_{j}$, ${\cal O}_{k}$ and derivatives.
Let us label the resulting double-trace primary scalar of dimension $\Delta_n=\Delta_j+\Delta_k +2n$ as $({\cal O}_j {\cal O}_k)_n$. 
The 3-point function $\langle K({\cal O}_j {\cal O}_k)_n(z,x) {\cal O}_{j}(y_1) {\cal O}_{k}(y_2)\rangle $ has a coefficient $c_{njk}$, so we have
\begin{equation}
\langle K({\cal O}_j {\cal O}_k)_n(z,x) {\cal O}_{j}(y_1) {\cal O}_{k}(y_2)\rangle =\frac{1}{(y_1 - y_2)^{2\Delta_j}}
\left[\frac{z}{z^2+(x-y_2)^2}\right]^{\Delta_k-\Delta_j} I^{(n)}(\chi)
\end{equation}
where  $I^{(n)}(\chi)$ is obtained from $I(\chi)$ by replacing $\Delta \rightarrow \Delta_{n}$, $\tilde{\gamma} \rightarrow c_{njk}$ (likewise
for $I_{\rm discont}^{(n)}(\chi)$).
We can then choose coefficients $a_{njk}$ such that
\begin{equation}
I_{\rm discont}(\chi)=-\frac{1}{N}\sum_{n=0}^{\infty}a_{njk}I_{\rm discont}^{(n)}(\chi).
\label{discan}
\end{equation}
Defining 
\begin{equation}
b_{njk}=a_{njk}(\Delta_{n}(\Delta_{n}-d)-\Delta(\Delta-d)),
\end{equation}
it was found in \cite{Kabat:2015swa} that (\ref{discan}) is satisfied (with the normalizations used here) provided
\begin{equation}
\frac{1}{N}\frac{b_{n12}c_{n12}}{2\Delta_{n}-d}=\frac{\lambda^{(s)}}{\pi^{d}}\frac{(-1)^{n}}{\Gamma(\Delta_{1}-\frac{d}{2})\Gamma(\Delta_{2}-\frac{d}{2})}
\frac{\Gamma(\Delta_{1}+n)\Gamma(\Delta_{2}+n)\Gamma(n+\Delta_{1}+\Delta_{2}-\frac{d}{2})}{\Gamma(n+1)\Gamma(2n+\Delta_{1}+\Delta_{2}-\frac{d}{2})}
\label{bncn}
\end{equation}
where
\begin{equation}
\lambda^{(s)}=\frac{\tilde{\gamma}}{\gamma_{F}(\Delta,\Delta_j,\Delta_k)}.
\end{equation}
This means that if we define a corrected bulk operator by
\begin{equation}
\phi(z,x)=\phi^{(0)}(z,x)+\frac{1}{N}\sum_{n=0}^{\infty} a_{njk} K({\cal O}_j {\cal O}_k)_n (z,x)
\label{bulkimp}
\end{equation}
it will give inside a 3-point function an expression obeying bulk causality.
In fact there are infinitely many solutions for $a_{njk}$ that satisfy (\ref{discan}). They are differ from the above by the substitution
\begin{equation}
b_{n12} \rightarrow b_{n12}+\sum_{l=1}^{\infty}\beta_{l}(M_{n}^{2}-m_{0}^2)^{l}b_{n12}
\end{equation}
Here $\beta_{l}$ are arbitrary.  They parametrize bulk field redefinitions that are compatible with the $1/N$ counting rules to this order.

Another useful result from \cite{Kabat:2015swa} is the sum
\begin{equation}
\frac{1}{N}\sum_{n=0}^{\infty} b_{njk}I^{(n)}(\chi)=\frac{\Gamma(\Delta_j)\Gamma(\Delta_k)}{\pi^{d}\Gamma(\Delta_j-\frac{d}{2})\Gamma(\Delta_k-\frac{d}{2})}\frac{\lambda^{(s)}}{\chi^{\Delta_j}}
\label{scalarsum}
\end{equation}
This implies that
\begin{equation}
\sum_{n=0}^{\infty}b_{njk}K({\cal O}_{j}{\cal O}_{k})_{n}(z,x)=\lambda^{(s)}K{\cal O}_{j}(z,x)K{\cal O}_{k}(z,x)
\end{equation}

\section{Bulk scalar -- scalar -- vector 3-point function}
We start with the 3-point function of a primary scalar of dimension $\Delta$, another primary scalar of dimension $\Delta_j$ and a non-conserved primary current of dimension $\Delta_{\rm v}$.
\begin{eqnarray}
&&\langle{\cal O}(x){\cal O}_{j}(y_1)j_{\mu}(y_2)\rangle =\left ( \frac{(y_1-y_2)_{\mu}}{(y_1-y_2)^2}- \frac{(x-y_2)_{\mu}}{(x-y_2)^2}\right ) \\
&& \qquad \times \frac{\gamma^{({\rm v})}}{(y_1-x)^{\Delta+\Delta_j-\Delta_{\rm v}+1}(y_1-y_2)^{\Delta_j+\Delta_{\rm v} -\Delta-1}(x-y_2)^{\Delta-\Delta_j+\Delta_{\rm v}-1}} \nonumber
\end{eqnarray}
This can be written as\footnote{correcting a typo in \cite{Kabat:2015swa}}
\begin{eqnarray}
\frac{\gamma^{({\rm v})}}{\Delta_{\rm v}+\Delta-\Delta_j -1}\left [(y_1 -y_2)^2\partial_{\mu}^{y_2}-2\Delta_{\rm v} (y_1 -y_2)_{\mu} \right ]\langle\Delta,\Delta_2 +1,\Delta_{\rm v}\rangle 
\label{sv3point1}
\end{eqnarray}
where $\langle\Delta,\Delta_2 +1,\Delta_{\rm v}\rangle $ is the canonical 3-point function of three scalars with unit coefficient.
We know that to promote ${\cal O}(x)$ to a local bulk operator 
we must add a tower of appropriately-smeared higher-dimension scalar primaries.
In appendix A this was done in (\ref{bulkimp}) to restore causality in a scalar 3-point function.
From (\ref{sv3point1}) we see that we can borrow this result to restore causality in this case as well.
Indeed adding smeared higher-dimension primaries with the appropriate coefficients results in a bulk operator obeying
\begin{eqnarray}
& & \langle(\nabla^2-m_{0}^2)\phi(z,x) {\cal O}_{j}(y_1) j_{\mu}(y_2)\rangle = \\
& & \qquad \frac{\gamma^{({\rm v})}}{\Delta_{\rm v}+\Delta-\Delta_j -1}\left [\frac{(y_1 -y_2)^2\partial_{\mu}^{y_2}-2\Delta_{\rm v} (y_1 -y_2)_{\mu}}{\gamma_{F}(\Delta,\Delta_j+1,\Delta_{\rm v})} \right]
\times(scalar \ result)\nonumber
\end{eqnarray}
where
\begin{equation}
(scalar \ result)=\frac{\Gamma(\Delta_{\rm v})\Gamma(\Delta_j+1)}{\pi^{d}\Gamma(\Delta_{\rm v}-\frac{d}{2})\Gamma(\Delta_j+1-\frac{d}{2})}\left (\frac{z}{(x-y_1)^2 +z^2}\right )^{\Delta_j +1}\left (\frac{z}{(x-y_2)^2 +z^2}\right )^{\Delta_{\rm v}}
\end{equation}
Thus after some computations we can identify 
\begin{equation}
(\nabla^2-m_{0}^2)\phi(z,x)=\frac{-\gamma^{({\rm v})}\Delta_{\rm v}}{\gamma_{F}(\Delta,\Delta_j+1,\Delta_{\rm v})(\Delta_j -\frac{d}{2})(\Delta_{\rm v}+\Delta-\Delta_j -1)}\partial^{M}K{\cal O}_{j}A_{M}(z,x)
\end{equation}
where $M=(z,x)$ and $A_{M}(z,x)$ is a bulk massive vector field expressed in terms of CFT operators by \cite{Kabat:2012hp}
\begin{eqnarray}
\label{mvecsmear1}
&&A^{(0)}_{z}(z,x)=\frac{1}{d-\Delta_{\rm v}-1}K_{\Delta_{\rm v}}(\partial^{\mu}j_{\mu})(z,x) \\
&&zA^{(0)}_{\mu}(z,x)=K_{\Delta_{\rm v}}j_{\mu}(z,x) +\frac{z}{2(\Delta_{\rm v}-\frac{d}{2}+1)}\frac{1}{d-\Delta_{\rm v}-1}K_{\Delta_{\rm v}+1}\partial_{\mu}(\partial^{\nu}j_{\nu})(z,x)\nonumber
\end{eqnarray}
We have adopted a normalization in which
\begin{eqnarray}
&&\frac{1}{d_{\Delta_{\rm v}}}\langle zA_{\nu}(x,z)j_{\mu}(y_1)\rangle =\frac{\Delta-1}{\Delta}\eta_{\mu \nu} \left(\frac{z}{(x-y_1)^2+z^2}\right)^{\Delta}-\frac{z^\Delta}{2\Delta(\Delta-1)}\partial^{x}_{\mu}\partial^{x}_{\nu}\left(\frac{1}{(x-y_1)^2+z^2}\right)^{\Delta-1}\nonumber\\
&&\frac{1}{d_{\Delta_{\rm v}}}\langle A_{z}(x,z)j_{\mu}(y_1)\rangle =\frac{1}{\Delta}\partial_{x_{\mu}} \left(\frac{z}{(x-y_1)^2+z^2}\right)^{\Delta}\nonumber \\
&&d_{\Delta_{\rm v}}=\frac{\Gamma(\Delta_{\rm v})}{\pi^{\frac{d}{2}}\Gamma(\Delta_{\rm v} -\frac{d}{2})}
\end{eqnarray}

\section{Bulk vector -- scalar -- scalar 3-point function}
We start with the 3-point function of a primary non-conserved current of dimension $\Delta_{\rm v}$ and two scalars with dimensions $\Delta_{j}=\Delta_k=\Delta$.
\begin{equation}
\langle j_{\mu}(x){\cal O}_{j}(y_1){\cal O}_{k}(y_2)\rangle =\left ( \frac{(y_1-x)_{\mu}}{(y_1-x)^2}- \frac{(y_2-x)_{\mu}}{(y_2-x)^2}\right)
\frac{\gamma^{({\rm v})}}{(y_1-y_2)^{2\Delta-\Delta_{\rm v}+1}(y_1-x)^{\Delta_{\rm v}-1}(y_2-x)^{\Delta_{\rm v}-1}}
\end{equation}
We can smear the non-conserved current as in (\ref{mvecsmear}) to get \cite{Kabat:2012av}
\begin{eqnarray}
\label{basicvec}
&&\langle A_{M}(x,z) {\cal O}_{j}(y_1) {\cal O}_{k}(y_2)\rangle =\frac{1}{2}\left ( \partial_{M}^{x}\ln \frac{(x-y_2)^2+z^2}{(x-y_1)^2+z^2}\right )\\
&& \qquad \times \frac{\frac{\gamma^{({\rm v})}}{2\Delta_{\rm v} -d}}{(y_1-y_2)^{2\Delta}(\chi-1)^{\frac{\Delta_{\rm v}-1}{2}}}F(\frac{\Delta_{\rm v}-1}{2},\frac{\Delta_{\rm v}-d+1}{2},\Delta_{\rm v}-\frac{d}{2}+1,\frac{1}{1-\chi}) \nonumber
\end{eqnarray}
The first line of (\ref{basicvec}) is independent of the operators involved, while the second line is (up to
a coefficient) the 3-point function of a smeared scalar operator of dimension $\Delta_{\rm v}$ with two other scalars of dimension $\Delta$ and $\Delta+1$.
To make a local massive vector in the bulk we need to correct the zeroth order definition by adding a tower of massive vectors \cite{Kabat:2012av}.
These operators are built from ${\cal O}_{j}$, ${\cal O}_{k}$ and derivatives, and we label the operator with dimension $\Delta_m=2\Delta+2m+1$ as
$({\cal O}_{j}{\cal O}_{k})^{m}_{\mu}$. This double-trace primary current has a non-trivial 3-point function with ${\cal O}_{j}$ ${\cal O}_{k}$, with a coefficient we
denote $c^{({\rm v})}_{mjk}.$\footnote{$c^{({\rm v})}_{mjk}$ can be computed from results in \cite{Fitzpatrick:2011dm}.} We label the zeroth order smearing of these
operators as in (\ref{mvecsmear}) by $A^{(m)}_{M}(z,x)$.  Then we have the 3-point function \cite{Kabat:2012hp, Kabat:2012av} 
 \begin{eqnarray}
\label{basicvec2}
&&\langle A^{(m)}_{M}(x,z) {\cal O}_{j}(y_1) {\cal O}_{k}(y_2)\rangle =\frac{1}{2}\left ( \partial_{M}^{x}\ln \frac{(x-y_2)^2+z^2}{(x-y_1)^2+z^2}\right )\\
&& \qquad \times \frac{\frac{c^{({\rm v})}_{mjk}}{2\Delta_m -d}}{(y_1-y_2)^{2\Delta}(\chi-1)^{\frac{\Delta_m-1}{2}}}F(\frac{\Delta_m-1}{2},\frac{\Delta_m-d+1}{2},\Delta_m-\frac{d}{2}+1,\frac{1}{1-\chi}) \nonumber
 \end{eqnarray}

From the relation to the scalar case noted above, we know that if we define a corrected bulk massive vector by
(here we are assuming $\gamma^{({\rm v})} \sim {1}/{N}$)
\begin{equation}
A_{M}(z,x)=A_{M}^{(0)}(z,x)+\frac{1}{N}\sum_{m} \frac{b^{({\rm v})}_{mjk}}{\Delta_{m}(\Delta_{m}-d)-\Delta_{\rm v} (\Delta_{\rm v} -d)}A_{M}^{(m)}(z,x)
\label{bulomass}
\end{equation}
with $b_{mjk}^{({\rm v})}$ obeying
\begin{eqnarray}
\label{bncnv}
&&\frac{1}{N}\frac{b_{mjk}^{({\rm v})}c^{({\rm v})}_{mjk}}{2\Delta_m -d}=\frac{\gamma^{({\rm v})}}{\gamma_{F}(\Delta_{\rm v},\Delta,\Delta+1)\pi^{d}}\frac{(-1)^{m}}{\Gamma(\Delta-\frac{d}{2})\Gamma(\Delta+1-\frac{d}{2})} \\
\nonumber
&& \hspace{3cm} \times \frac{\Gamma(\Delta+m)\Gamma(\Delta+1+m)\Gamma(m+2\Delta+1-\frac{d}{2})}{\Gamma(m+1)\Gamma(2m+2\Delta+1-\frac{d}{2})}
\end{eqnarray}
then (\ref{bulomass}) will obey locality inside 3-point functions.
 
\subsection{Vector sum}
In this section we wish to show the following. We start with the infinite  tower of primary non-conserved currents $({\cal O}_{j}{\cal O}_{k})^{m}_{\mu}$  defined
above, with dimension $\Delta_m=2\Delta+2m+1$. We then smear each one appropriately as in (\ref{mvecsmear}) and label it $A^{(m)}_{M}(z,x)=K({\cal O}_{j}{\cal O}_{k})^{m}_{M}(z,x)$, where $M=(z,x)$. Then there are coefficients $\tilde{b}^{({\rm v})}_{mjk}$ which obey (to leading order in $1/N$) the following.
 \begin{equation}
(2\Delta -d) \sum \tilde{b}^{({\rm v})}_{mjk}(K{\cal O}_{j}{\cal O}_{k})^{m}_{M}(z,x)=\phi_{k}(z,x)\partial_{M}\phi_{j}(z,x)-\phi_{j}(z,x)\partial_{M}\phi_{k}(z,x)
\label{vss1}
 \end{equation}

To show this we insert both sides of (\ref{vss1}) into a 3-point function with ${\cal O}_{j}(y_1)$ and 
${\cal O}_{k}(y_2)$.\footnote{Here we treat the case where $M=\mu$, a coordinate parallel to the boundary.  A similar argument holds
for the $z$ coordinate.}
On the right we get 
\begin{eqnarray}
\label{vss1r}
&&\langle(\phi_{k}\partial_{\mu}\phi_{j}-\phi_{j}\partial_{\mu}\phi_{k})(z,x) {\cal O}_{j}(y_1){\cal O}_{k}(y_2)\rangle =
2\Delta \frac{\Gamma^{2}(\Delta)}{\pi^{d}\Gamma^{2}(\Delta-\frac{d}{2})} \\
&& \qquad \times \left (\frac{z}{(x-y_1)^2 +z^2}\right )^{\Delta} \left (\frac{z}{(x-y_2)^2 +z^2}\right )^{\Delta} \left (\frac{(x-y_2)_{\mu}}{(x-y_2)^2+z^2}-\frac{(x-y_1)_{\mu}}{(x-y_1)^2+z^2} \right )\nonumber
\end{eqnarray}

On the left we can use the result (\ref{basicvec2}) for the 3-point function of a primary non-conserved current of dimension $\Delta_m=2\Delta+2m+1$ with
two scalars of dimension $\Delta_j=\Delta_k=\Delta$.
From the scalar case (\ref{scalarsum}) we know that for  $\tilde{b}_{mjk}^{({\rm v})}$ such that
\begin{equation}
\tilde{b}_{mjk}^{({\rm v})}=\frac{1}{N}b_{mjk}^{({\rm v})}\frac{\gamma_{F}(\Delta_{\rm v},\Delta,\Delta+1)}{\gamma^{({\rm v})}}
\end{equation}
we have
\begin{eqnarray}
\nonumber
&&\frac {\tilde{b}_{mjk}^{({\rm v})}c^{({\rm v})}_{mjk}}{2\Delta_m -d}\frac{1}{(\chi-1)^{\frac{\Delta_m-1}{2}}}F(\frac{\Delta_m-1}{2},\frac{\Delta_m-d+1}{2},\Delta_m-\frac{d}{2}+1,\frac{1}{1-\chi}) \\
&& \qquad\qquad =\frac{\Delta \Gamma^{2}(\Delta)}{\pi^{d}(\Delta-\frac{d}{2})\Gamma^{2}(\Delta-\frac{d}{2})}\frac{1}{\chi^{\Delta}}
\end{eqnarray}
 So the left-hand side of (\ref{vss1}) inserted into the 3-point function gives
\begin{equation}
\frac{\Delta\Gamma^{2}(\Delta)}{\pi^{d}\Gamma^{2}(\Delta-\frac{d}{2})} \frac{1}{(y_1-y_2)^{2\Delta}}\left ( \partial_{M}^{x}\ln \frac{(x-y_2)^2+z^2}{(x-y_1)^2+z^2}\right )\frac{1}{\chi^{\Delta}}
\end{equation}
which exactly agrees with (\ref{vss1r}).

\providecommand{\href}[2]{#2}\begingroup\raggedright\endgroup

\end{document}